\documentclass[%
12pt,
 superscriptaddress,
 amsmath,amssymb,
 aps,UTF8,
]{revtex4-1}

\usepackage{graphicx}
\usepackage{dcolumn}
\usepackage{bm}
\usepackage{xcolor}
\usepackage{url}
\usepackage{float}
\usepackage{tabularx}
\usepackage{lipsum}
\usepackage{array}

\usepackage{booktabs}

\usepackage{subfigure}
\usepackage{multirow}

\begin{document}


\title{Deriving the QCD evolution equations under the Abelian decomposition scheme}

\email{These authors contributed equally: Yirui Yang and Wei Kou.}

\author{Yirui Yang}
\affiliation{Institute of Modern Physics, Chinese Academy of Sciences, Lanzhou 730000, China}
\affiliation{School of Physics, Peking University, Beijing 100817, China}
\author{Wei Kou}
\email{kouwei@impcas.ac.cn}
\affiliation{Institute of Modern Physics, Chinese Academy of Sciences, Lanzhou 730000, China}
\affiliation{School of Nuclear Science and Technology, University of Chinese Academy of Sciences, Beijing 100049, China}
\author{Xiaopeng Wang}
\email{wangxiaopeng@impcas.ac.cn}
\affiliation{Institute of Modern Physics, Chinese Academy of Sciences, Lanzhou 730000, China}
\affiliation{School of Nuclear Science and Technology, University of Chinese Academy of Sciences, Beijing 100049, China}
\affiliation{Lanzhou University, Lanzhou 730000, China}
\author{Yanbing Cai}
\email{yanbingcai@mail.gufe.edu.cn (Correspongding author)}
\affiliation{Guizhou Key Laboratory in Physics and Related Areas, Guizhou University of Finance and Economics, Guiyang 550025, China}
\affiliation{Southern Center for Nuclear-Science Theory (SCNT), Institute of Modern Physics, Chinese Academy of Sciences, Huizhou 516000, China}
\author{Xurong Chen}
\email{xchen@impcas.ac.cn (Correspongding author)}
\affiliation{Institute of Modern Physics, Chinese Academy of Sciences, Lanzhou 730000, China}
\affiliation{School of Nuclear Science and Technology, University of Chinese Academy of Sciences, Beijing 100049, China}



\begin{abstract}
The Abelian decomposition of QCD reveals two types of gluons: color-neutral ``neurons" and color-carrying ``chromons". This classification does not alter the overall properties of QCD, but the investigation of different types of gluon dynamics is necessary. By employing the Cho-Duan-Ge decomposition theorem, we have derived dynamic evolution equations for two types of gluons by using the time-ordered perturbation theory. We propose that the new equations are compatible with the DGLAP equations, requiring only the separate contributions of neurons and chromons to be summed. Surprisingly, with the evolution to high $Q^2$, the ratio of the number of chromons to neurons is approximately 3:1 in small-$x$ region regardless of the inputs at evolution starting point. The new gluon dynamic equations reevaluate the gluon distribution functions and allow for a elaborate inverstigation of the distinct contributions of gluons in high-energy collisions.

\end{abstract}

\maketitle


\section{Introduction}
\label{sec:intro}

The non-Abelian Yang-Mills theory reveals that the color gauge field (gluon) confines quarks within hadrons, which contradicts the quark model's depiction of quarks composing hadrons. The quark model lacks a description of gluons, raising the question of how quarks bind together. Do gluons exist in only one type, and must they adhere to color symmetry, reflecting the question of whether non-Abelian color gauge theories can be simplified for understanding? The Abelian decomposition theory of QCD \cite{Cho:1979nv,Duan:1979ucg,Cho:1980nx,Cho:1981ww} posits the possibility of disentangling the substructure of the non-Abelian gauge potential into an Abelian component, denoted as restricted QCD (RCD). RCD delineates the Abelian dynamics of QCD while upholding non-Abelian color gauge symmetry. After nearly 40 years of refinement, the Abelian decomposition has developed a systematic theoretical framework and is now known as the Cho-Duan-Ge (CDG) decomposition or the Cho-Faddeev-Niemi (CFN) decomposition \cite{Faddeev:1998eq,Faddeev:1998yz,Shabanov:1999uv,Shabanov:1999xy,Gies:2001hk,Zucchini:2003cy,Kondo:2014sta}.

One of the motivations behind the Abelian decomposition lies in the foundations of group theory. Within group theory, it is described that the color gauge group contains an Abelian subgroup, characterized by its generators being diagonal matrices that correspond to the color-neutral gauge potentials. The remaining non-diagonal generators must necessarily carry color information. This implies the existence of two types of gauge fields (gluons): those that are color-neutral and those that carry color. Another motivation is linked to QCD confinement. Color confinement is a complex subject, and the concepts of monopole condensation \cite{Nambu:1974zg,Mandelstam:1974pi,Polyakov:1976fu,Cho:1980nx,Cho:1981ww} and Abelian dominance \cite{tHooft:1981bkw,Cho:1999ar} can be considered as candidate explanations for the mechanism of confinement. These two areas have propelled research into the Abelian decomposition of QCD. It is necessary to emphasize the possible explanation of QCD confinement that is reflected in the Abelian decomposition theory. Since gluons can be divided into two types of dynamical gluons -- neuron and chromon, as well as the existence of topological structure monopole, the carriers of confinement effects can be analyzed class by class. Firstly, the color-neutral neuron does not participate in color interactions, and therefore is irrelevant to confinement. Secondly, the chromon itself carries color charge and is the object of confinement, so it cannot be the cause of confinement. Finally, only the monopole, as a topological part of the gauge potential, is the "source" behind confinement. Therefore, exploring the dynamics of these gauge potentials and the topological structure is one of the possible angles for studying confinement.

The Abelian decomposition scheme involves decomposing the non-Abelian gauge potential into two components: the restricted Abelian part, which retains the full non-Abelian gauge symmetry, and the gauge-covariant valence part, which describes colored gluons. Furthermore, the restricted part can be further subdivided into the non-topological Maxwell structure and the topological monopole part \cite{Cho:1979nv,Duan:1979ucg,Cho:1980nx,Cho:1981ww}. The specific formal expressions for the decomposition and relevant physical insights can be found in \cite{Cho:2022rib}. The experimental demonstration of the existence of two types of gluons (excluding topological monopoles for the time being) is an urgent matter. Despite the interdisciplinary integration in the CDG theory, particle physics experiments have not yet definitively indicated the discovery of signals for two types of gluons. Particle physicists often utilize the final state jet distributions produced in hadron or heavy-ion high-energy collision experiments to study the hadronization process and investigate the quark and gluon structure functions within hadrons. It is suggested in \cite{Cho:2019mde,Cho:2022rib} to analyze the existence of two types of gluons and their respective contributions to jet emissions from the perspective of gluon jets.

The experimental consideration of the existence of two types of gluons is rather challenging. Fortunately, the bridge between QCD theory and experiments, the evolution equations of partons, holds greater practical significance. The theoretical explanation of high-energy collision experiments requires parton distribution functions at different scales, which are ensured by the evolution equations of QCD. Classical QCD evolution equations, such as the Dokshitzer-Gribov-Lipatov-Altarelli-Parisi (DGLAP) equation \cite{Dokshitzer:1977sg,Gribov:1972ri,Gribov:1972rt,Lipatov:1974qm,Altarelli:1977zs} describing the high virtuality $Q^2$ probe region and the Balitsky-Fadin-Kuraev-Lipatov (BFKL) dynamic equation \cite{Kovchegov:2012mbw} for the high-energy limit at small $x$, have provided precise predictions for high-energy physics experiments. The Abelian decomposition yields two types of gluons involved in the dynamical evolution, making it particularly important to obtain evolution equations for these two types of gluons. It is reasonable to hypothesize that the total contributions from the evolution equations of the two types of gluons should be consistent with the complete non-Abelian gluon evolution equation. In principle, we first consider the evolution equations for the gauge potentials in the $Q^2$ direction, namely the QCD dynamical evolution equations of neurons and chromons.

In this study, we first provide a brief overview of the origins of the two types of gluons and their corresponding physical meanings. Then, we present the relevant splitting functions for neurons and chromons, which are similar to the derivation process of the DGLAP equation. In the main text, we directly present the dynamical evolution equations for the two types of gluons and the quarks involved. Additionally, we present the results of numerical simulations, obtaining the parton distribution functions for neurons and chromons, which can be used as inputs for computing cross-sections and other physical quantities in high-energy collision processes.

\section{Abelian decomposition and Feynman Rules}
\label{sec:formalism}
In this section, we provide a simple overview of the Abelian decomposition following the main results in \cite{Cho:2022rib}. In order to demonstrate the existence of two types of gluons in QCD, we first review the Abelian projection of the SU(2) group to obtain the restricted part and the gauge-covariant part of the non-Abelian gauge potential. We choose an arbitrary basis of the SU(2) group as $\hat{n}=\hat{n}_1,\ \hat{n}_2,\ \hat{n}_3$ and select $\hat{n}$ as the Abelian direction. Projecting the non-Abelian gauge field $\vec{A}_\mu$ onto the direction of $\hat{n}$ yields the restricted gauge field, denoted as $\hat{A}_\mu$,
\begin{equation}
	\begin{aligned}
		D_{\mu}\hat{n}& =(\partial_\mu+g\vec{A}_\mu\times)\hat{n}=0,  \\
		\vec{A}_{\mu}\to\hat{A}_{\mu}& =A_\mu\hat{n}-\frac1g\hat{n}\times\partial_\mu\hat{n}=\tilde{A}_\mu+\tilde{C}_\mu,  \\
		\tilde{A}_{\mu}& =A_\mu\hat{n},\quad\tilde{C}_\mu=-\frac1g\hat{n}\times\partial_\mu\hat{n}.
	\end{aligned}
	\label{eq:CDG}
\end{equation}
The introduction of the vector $\hat{n}$ results in the restricted part $\hat{A}_\mu$ being characterized by a non-topological potential $\tilde{A}_\mu$ and a topological potential $\tilde{C}_\mu$. In order to restore the complete gauge symmetry of SU(2), the gauge-covariant colored potential $\vec{X}_\mu$ is introduced \cite{Cho:1979nv,Duan:1979ucg,Cho:1980nx,Cho:1981ww},
\begin{equation}
	\begin{aligned}
		\vec{A}_{\mu} =\hat{A}_\mu+\vec{X}_\mu, \ \ \
		\hat{n}\cdot\vec{X}_{\mu}=0.
	\end{aligned}
	\label{eq:colored}
\end{equation}
In this way, the QCD Lagrangian under the complete SU(3) gauge after the decomposition is expressed as,
\begin{equation}
	\begin{aligned}\mathcal{L}^\mathrm{SU(3)}_{\mathrm{QCD}}
    &=-\frac14\hat{F}_{\mu\nu}^2-\frac14(\hat{D}_{\mu}\vec{X}_{\nu}-\hat{D}_{\nu}\vec{X}_{\mu})^2\\
    &-\frac g2\hat{F}_{\mu\nu}\cdot(\vec{X}_{\mu}\times\vec{X}_{\nu})-\frac{g^2}4(\vec{X}_{\mu}\times\vec{X}_{\nu})^2\\
    &-\frac g2(\hat{D}_\mu\vec{X}_\nu-\hat{D}_\nu\vec{X}_\mu)\cdot(\vec{X}_\mu\times\vec{X}_\nu),
    \end{aligned}
	\label{eq:SU3QCD}
\end{equation}
with the decomposed gauge field strength tensor $\hat{F}_{\mu\nu}=\partial_\mu\hat{A}_{\nu}-\partial_\nu\hat{A}_{\mu}$. The Abelian decomposition reveals that under the SU(3) framework, the QCD Lagrangian can be partitioned into the restricted component and the chromon component with colored sources.

In addition to the gauge fields, incorporating the quark fields into the Lagrangian is essential. The original fermion current covariant derivative term undergoes modifications after the Abelian decomposition, leading to the following expression for the quark field component,
\begin{equation}
\label{eq:quark}
\begin{aligned}\mathcal{L}_Q^I&=\bar{\Psi}(i\gamma^\mu D_\mu-m)\Psi\\
&=\bar{\Psi}(i\gamma^\mu\hat{D}_\mu-m)\Psi+\frac{g}{2}\vec{X}_\mu\cdot\bar{\Psi}(\gamma^\mu\vec{t})\Psi,
\end{aligned}
\end{equation}
with the covariant derivative term $\hat{D}_\mu=\partial_\mu+\frac g{2i}\vec{t}\cdot\hat{A}_\mu$. Here $m$ is quark mass.

Now, the Abelian decomposition of QCD in the SU(3) case has been completed, and the primary task is to quantize the restricted field theory part to obtain the complete Feynman rules. We employ the path integral method to discuss the quantization results of each part of the Lagrangian and obtain the corresponding Feynman rules. First, we present the classical imagery of gluons decomposing into the neuron and chromon under the Abelian decomposition. In our representation, the classical gluons is represented by the spring-like line, while the color-neutral neuron (N) is represented by wavy line (bearing a striking resemblance to the properties of photon in QED). Additionally, we illustrate the colored chromon (C) as solid lines without arrows, to distinguish them from the arrowed quark (Q) lines. The Feynman diagrams resulting from the decomposition correspond one-to-one with the interaction terms of the SU(3) Lagrangian described earlier.

In order to obtain the full interaction terms corresponding to Feynman rules, we first need to extract the Abelian decomposed version of the Lagrangian for the quark-gluon vertex and the three-gluon vertex that will be used, and then simplify it as follows (the Abelian decomposition results excluding the free gauge field, fermion field, and the four-gluon vertex.):
\begin{equation}
	\mathcal{L}_{\text{int}} = \mathcal{L}_{3V}^{I} + \mathcal{L}_{Q}^{I},
	\label{eq:interaction term}
\end{equation}
where the subscript $3V$ represents the three-gluons vertex. It is evident that the quark-gluon vertex should have two cases after the Abelian decomposition, namely the vertices of the two types of gluons with the quark. Based on the Abelian decomposition, we can express the classical QCD's three-gluon vertex in terms of the two types of gluon fields,
\begin{equation}
	\begin{aligned}
		\mathcal{L}_{3V}^I&= -\frac{g}{2}(\partial_\mu \vec{X}_\nu-\partial_\nu \vec{X}_\mu)(\hat{A}_\mu\times\vec{X}_\nu -\hat{A}_\nu\times\vec{X}_\mu)\\
  &-\frac{g}{2}(\partial_\mu \vec{X}_\nu-\partial_\nu \vec{X}_\mu)(\vec{X}_\mu\times\vec{X}_\nu)-\frac{g}{2}(\partial_\mu \hat{A}_\nu-\partial_\nu \hat{A}_\mu)(\vec{X}_\mu\times\vec{X}_\nu)\\
          &-\frac{g^2}{4}(\hat{A}_\mu\times\vec{X}_\nu -\hat{A}_\nu\times\vec{X}_\mu)^2-\frac{g^2}{2}(\hat{A}_\mu\times\vec{X}_\nu -\hat{A}_\nu\times\vec{X}_\mu)(\vec{X}_\mu\times\vec{X}_\nu)-\frac{g^2}{4}(\vec{X}_\mu\times\vec{X}_\nu)^2\\
		&=\mathcal{L}_{NCC}^I+\mathcal{L}_{CCC}^I,
	\end{aligned}
	\label{eq:3-g vertex}
\end{equation}
where terms $\mathcal{L}_{NCC}^I$ and $\mathcal{L}_{CCC}^I$ respectively represent the two summation contents in the above equation. We emphasize that due to the requirement of color conservation, the three-gluon vertex does not include the combinations such as NNN and NNC \cite{Cho:2022rib}.

\begin{figure*}[]
	\centering
		\includegraphics[width=0.7\textwidth]{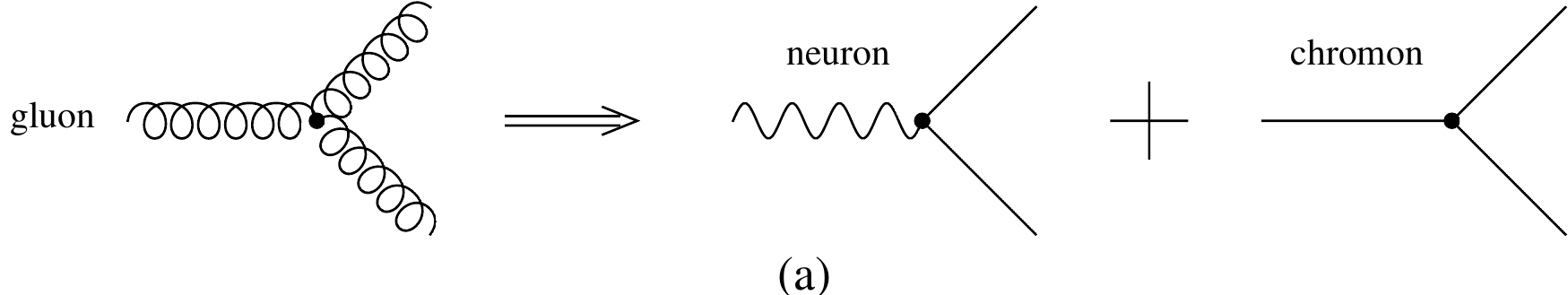}
		\label{fig-3g}
\vspace{0.5cm}
		\includegraphics[width=0.7\textwidth]{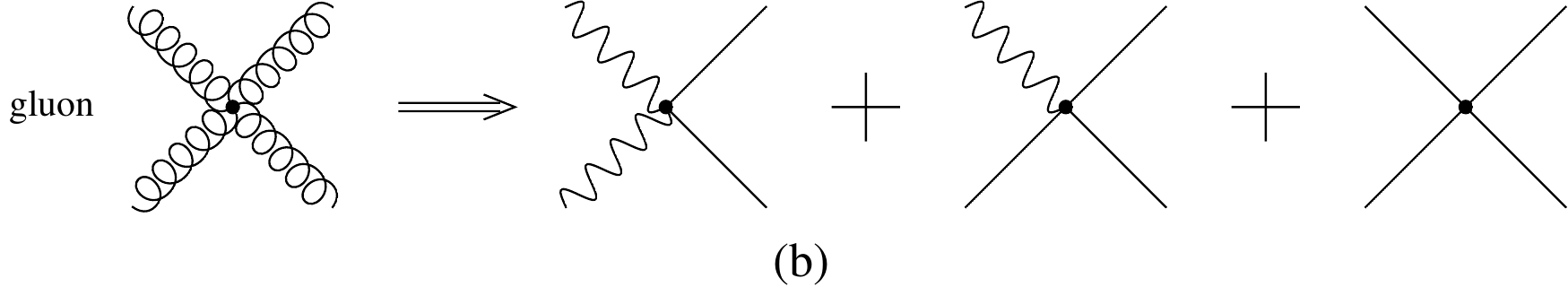}
		\label{fig-4g}
\vspace{0.5cm}
		\includegraphics[width=0.7\textwidth]{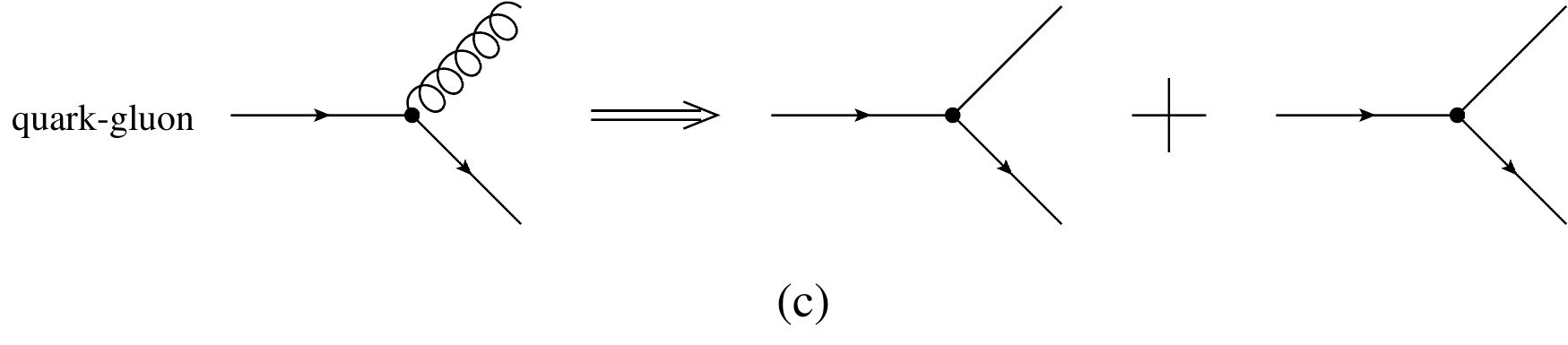}
		\label{fig-qg}
	\caption{The decomposition of the Feynman diagrams in SU(3) QCD: (a) the three-gluon, (b) four-gluon, (c) quark-gloun vertices.}
	\label{fig-guifan-2}
\end{figure*}

The Feynman diagrams corresponding to the interaction vertices are now clear (shown in Fig. \ref{fig-guifan-2}), and the Feynman rules can be obtained from the path integral quantization scheme, which we summarize in Table \ref{tab:FRs}. The Feynman rules for the vertices are crucial for deriving the splitting functions (parton evolution probabilities) in our QCD evolution equations. It is important to emphasize that we are currently only considering the leading-order (LO) approximation of DGLAP-like evolution, and therefore the four-gluon vertex is not within the scope of our consideration.
\begin{table}[]
	\centering
		\caption{Feynman rules for three-vertices under Abelian decomposition.}
		\resizebox{\columnwidth}{!}{
	\begin{tabular}{ c  c  c  }
		\hline
		Vertices& Feynman diagrams & Feynman Rules  \\ \hline
		QQN &
		\begin{minipage}[b]{0.3\columnwidth}
			\centering
			\raisebox{-.3\height}{\includegraphics[width=0.6\linewidth]{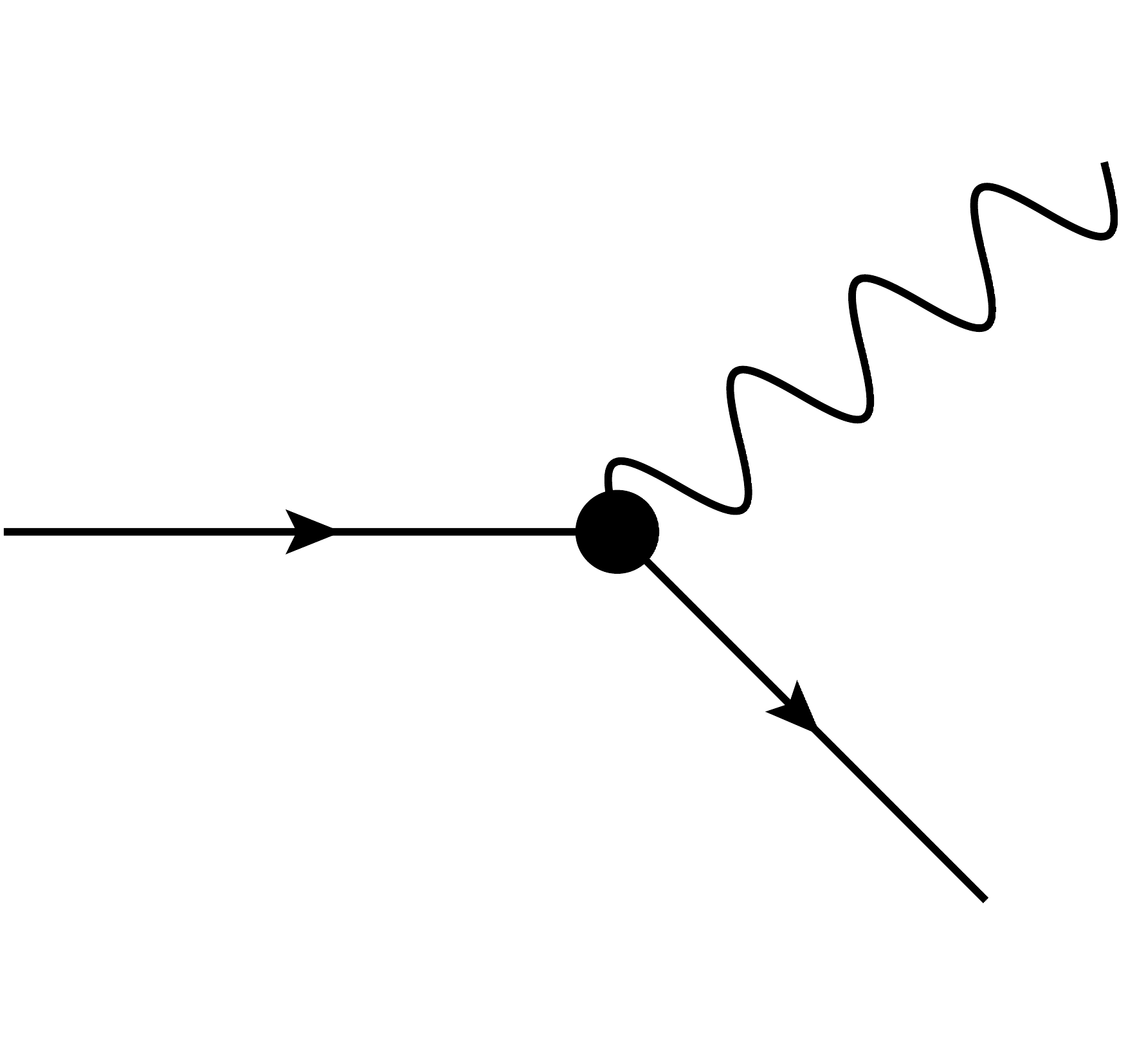}}
		\end{minipage}
		& $g\gamma_\mu(\frac{t^{a}}{2})_{ij},\, a\neq\,3,8 $
		\\ \hline
		QQC &
		\begin{minipage}[b]{0.3\columnwidth}
			\centering
			\raisebox{-.3\height}{\includegraphics[width=0.6\linewidth]{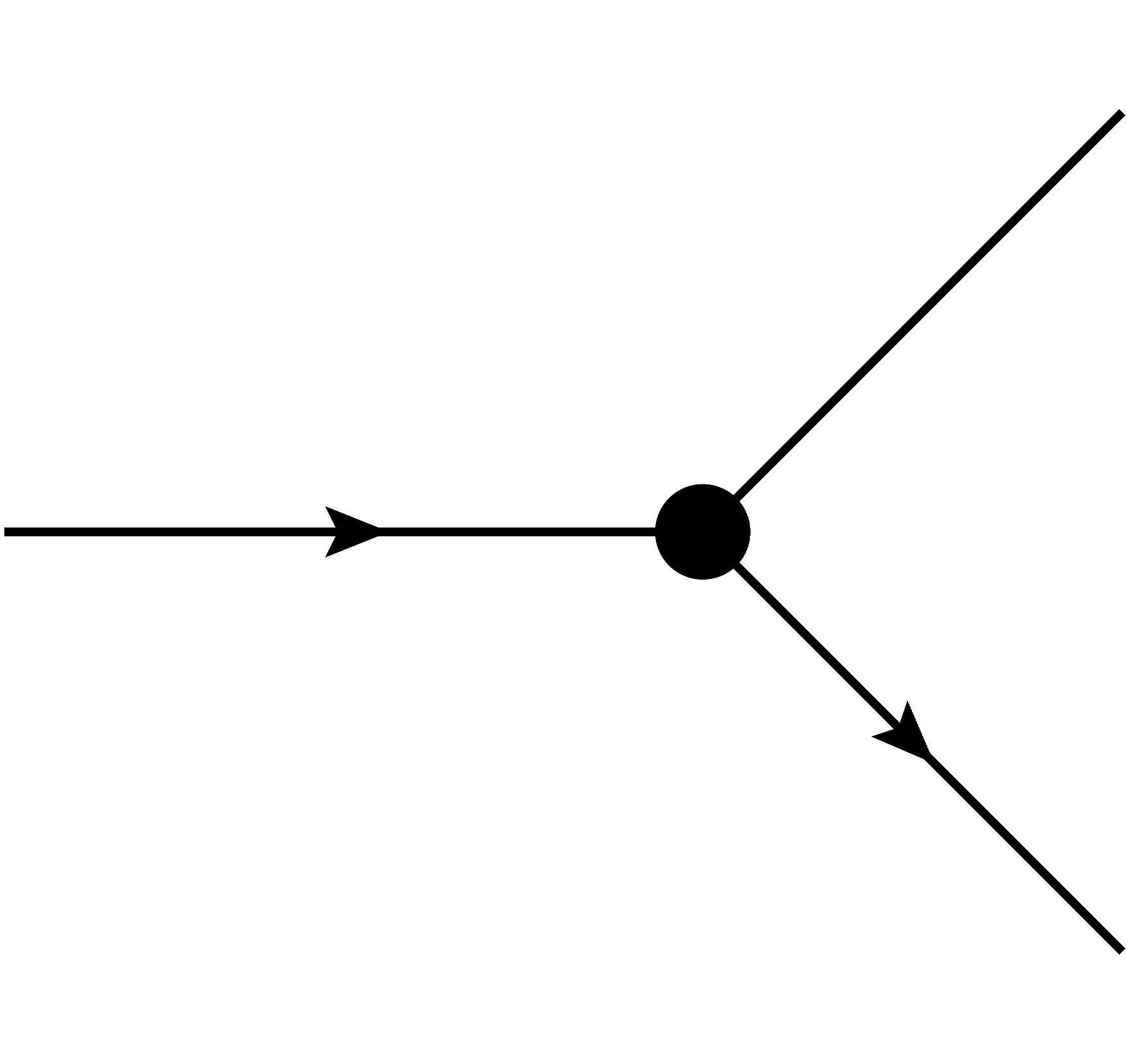}}
		\end{minipage}
		& $g\gamma_\mu(\frac{t^{\hat{a}}}{2})_{ij},\ \hat{a}=\,3,8 $
		\\ \hline
		NCC &
		\begin{minipage}[b]{0.3\columnwidth}
			\centering
			\raisebox{-.3\height}{\includegraphics[width=0.6\linewidth]{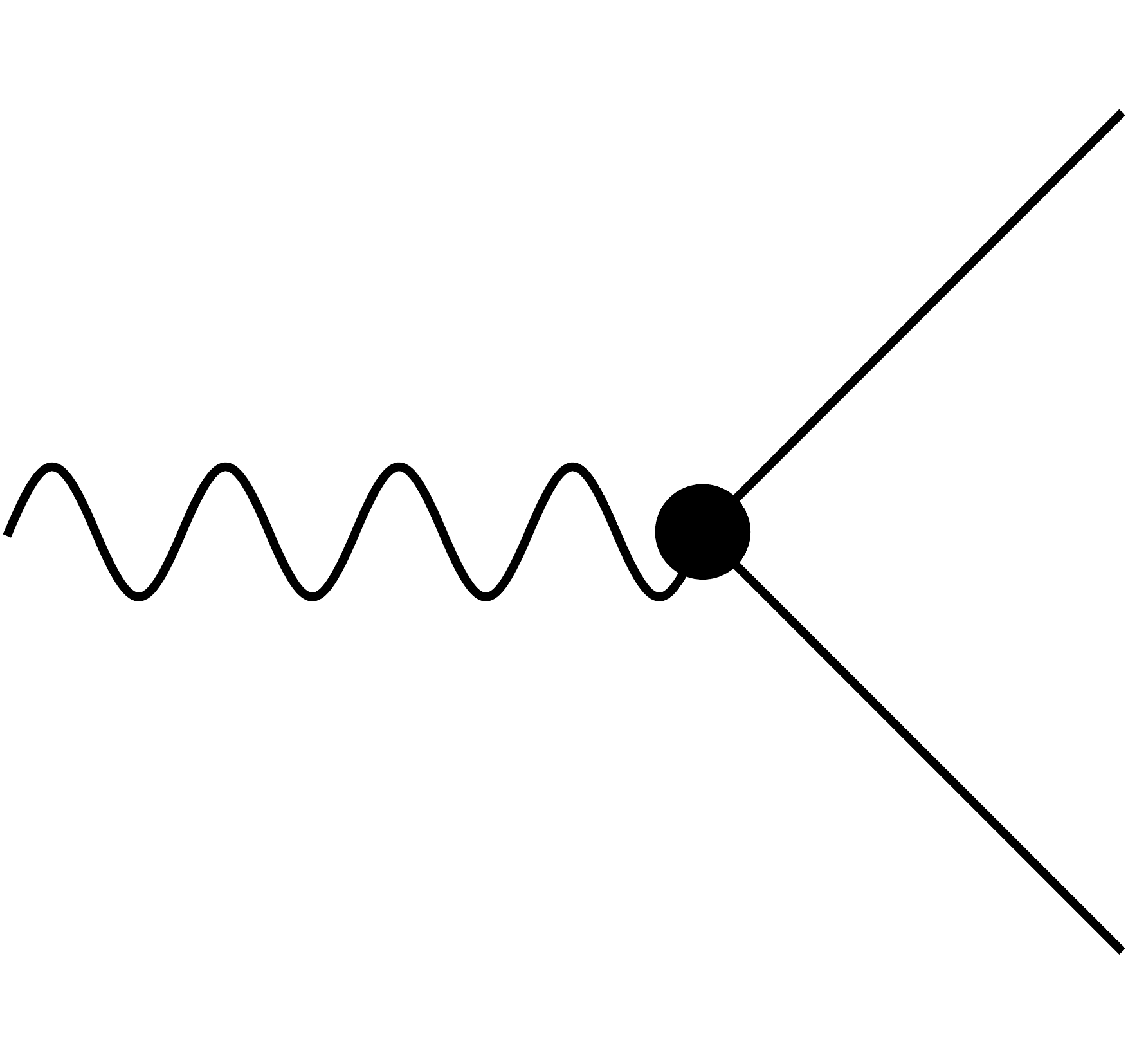}}
		\end{minipage}
		& $gf^{a\hat{b}c}\left( (g_{\mu \nu}(p_1-p_2)_\rho+g_{\nu\rho}(p_2-p_3)_\mu+g_{\rho\mu}(p_3-p_1)_\nu\right)\,a,c\neq\,3,8,\,\hat{b}=\,3,8$
		\\ \hline
		CCC &
		\begin{minipage}[b]{0.3\columnwidth}
			\centering
			\raisebox{-.3\height}{\includegraphics[width=0.6\linewidth]{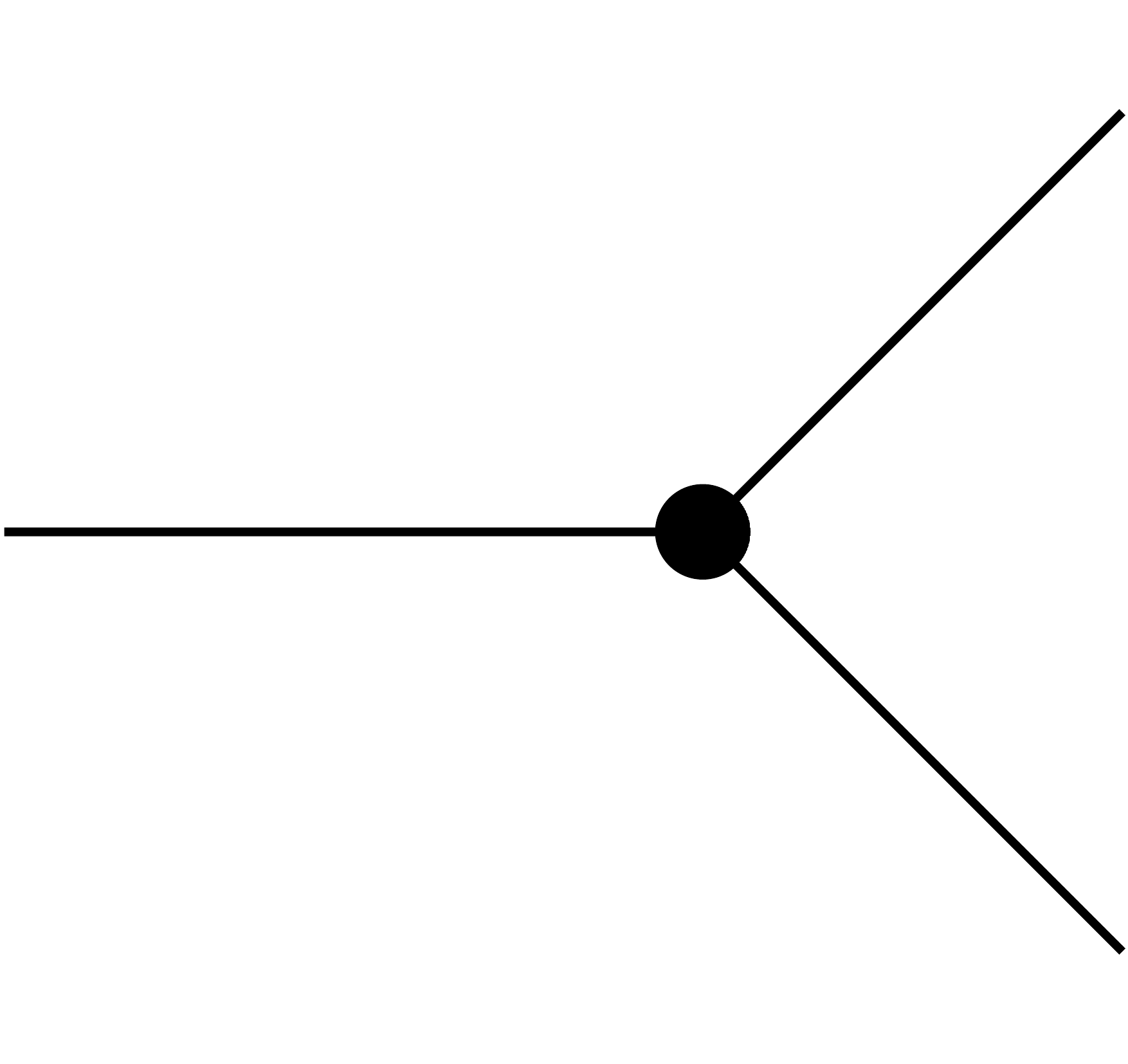}}
		\end{minipage}
		& $-gf^{abc}\left[g_{\mu\nu}(p_1-p_2)_\rho+g_{\nu\rho}(p_2-p_3)_\mu+g_{\rho\mu}(p_3-p_1)_\nu\right],\, a,b,c \neq \,3,8 $
		\\ \hline
	\end{tabular}}
	\label{tab:FRs}
\end{table}

\section{Evolution of two types of gluons}
\label{sec:new_evolution}
The original QCD DGLAP equation was obtained independently by Altarelli and Parisi \cite{Altarelli:1977zs} and Dokshitzer \cite{Dokshitzer:1977sg}, as follows:
\begin{widetext}
	\begin{equation}
		\begin{gathered}
			\frac{dq^i(x,t)}{dt} \begin{aligned}&=\frac{\alpha(t)}{2\pi}\int_x^1\frac{dy}y\left[\sum_{j=1}^{2f}q^j(y,t)P_{q^iq^j}(\frac xy)+G(y,t)P_{q^iG}(\frac xy)\right],\end{aligned} \\
			\frac{dG(x,t)}{dt} =\frac{\alpha(t)}{2\pi}\int_x^1\frac{dy}y\left[\sum_{j=1}^{2f}q^j(y,t)P_{Gq^j}(\frac xy)+G(y,t)P_{GG}(\frac xy)\right],
		\end{gathered}
		\label{eq:DGLAP}
	\end{equation}
	\end{widetext}
where $x$ denotes the momentum fraction of partons in the proton, $t=\ln(Q^2/Q_0^2)$ ($Q^2$ is the virtuality of the photon). In Eq. \ref{eq:DGLAP}, $\alpha(t)$ is the running coupling \cite{Workman:2022ynf}. The $q^i(x,t)$ and $G(x,t)$ to respectively denote the distribution functions of quarks of flavor $i$ and gluons. The $P_{ij}(z)$ represents the evolution kernel, also known as the splitting function, which describes the probability amplitude for parton $j$ to transition into parton $i$. Form the splitting functions in Eq. (\ref{eq:DGLAP}), we can find the transformation behavior between partons in the original QCD evolution, such as quark to quark, gluon to quark, quark to gluon, and gluon to gluon.

In Ref. \cite{Altarelli:1977zs}, the evolution equation was derived basing on the time-ordered perturbation theory (TOPT), which provides a more detailed description of the intermediate processes. Our derivation is based on TOPT. The QCD Feynman rules for the Abelian decomposition reveal that the three-gluon vertex and quark-gluon vertex undergo changes due to the inclusion of neuron and chromon. The splitting functions from the original evolution equations are no longer applicable, necessitating the derivation of new splitting functions. In addition to the pre-existing quark-to-quark splitting process ($P_{q^jq^i}$), the gluon-to-quark splitting process is divided into neuron-to-quark ($P_{qN}$) and chromon-to-quark ($P_{qC}$) processes. Similarly, the quark-to-gluon process is divided into neuron and chromon ($P_{Nq}$ and $P_{Cq}$), as allowed by the Feynman rules. Furthermore, the three-gluon vertex complicates the gluon-to-gluon splitting process. Apart from the absence of a three-neuron vertex under the color conservation requirement ($P_{NN}=0$), a neuron can convert into a chromon ($P_{CN}$), a chromon can convert into a new chromon ($P_{CC}$), and a chromon can convert into a neuron ($P_{NC}$).

To obtain the new QCD evolution equations in the Abelian decomposition, we first compute the corresponding splitting functions. For this purpose, we will first study the case where $z<1$, and then the singularity of $\delta$-function at $z=1$. Specifically, we need to calculate the probability of finding a particle $B$ in particle $A$ with momentum fraction $z$ in the infinite momentum frame. In the case of lowest order of $\alpha$, we have
	\begin{equation}
		\label{eq-ten5.5}
		d \mathcal{P}_{B A}(z) d z=\frac{\alpha}{2 \pi} \mathcal{P}_{B A}(z) d t\left(t=\ln \frac{Q^{2}}{Q_{0}^{2}}\right).
	\end{equation}
	
\begin{figure}[H]
		\centering
		\includegraphics[width=0.7\textwidth]{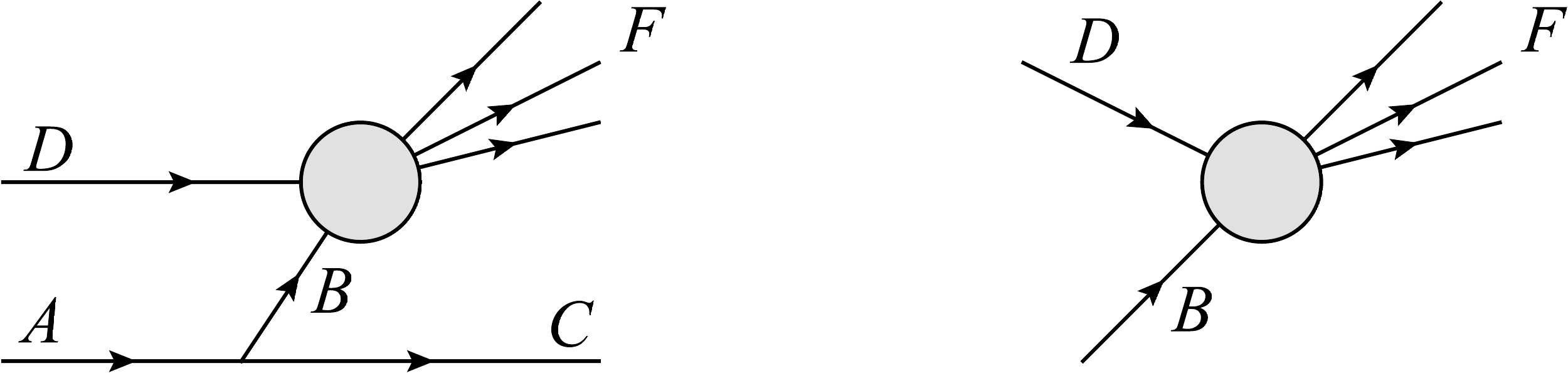}
		\caption{Scatterings of the (left) process $A+D\to{}C+F$ and (right) sub-process $B+D\to{}F$, where $D$ is a given particle and $F$ any final state.}
		\label{fig-tentopt0}
	\end{figure}

	As is shown in Fig.~\ref{fig-tentopt0}, a bare vertex $C$ between $A$ and $B$ is introduced, which is the intermediate state between $A$ and $B$. By analyzing the scattering cross sections of the two processes in Fig.~\ref{fig-tentopt0}, the corresponding probabilities can be obtained.
	
	The corresponding cross-sections shown in  Fig.~\ref{fig-tentopt0} can be easily computed and is therefore given by~\cite{Altarelli:1977zs}
	\begin{align}
		d \sigma_{a} =&\frac{1}{8 E_{A} E_{D}} \frac{\left|M_{A \rightarrow B+C}\right|^{2}\left|M_{B+D \rightarrow F}\right|^{2}}{\left(2 E_{B}\right)^{2}\left(E_{B}+E_{C}-E_{A}\right)^{2}} \nonumber\\
		& \times(2 \pi)^{4} \delta^{4}\left(k_{A}+k_{D}-k_{C}-k_{F}\right) \frac{d^{3} k_{C}}{(2 \pi)^{3}\left(2 E_{C}\right)} \prod_{F} \frac{d^{3} k_{F}}{(2 \pi)^{3}\left(2 E_{F}\right)},\label{eq-ten1}\\
		d \sigma_{b}=&\frac{g^{2}}{8 E_{B} E_{D}}\left|M_{B+D \rightarrow F}\right|^{2}(2 \pi)^{4} \delta^{4}\left(k_{B}+k_{D}-k_{F}\right) \prod_{F} \frac{d^{3} k_{F}}{(2 \pi)^{3}\left(2 E_{F}\right)}.\label{eq-ten2}
	\end{align}
	The two processes are related through
	\begin{equation}
		d \sigma_{a}=d {\cal{P}}_{BA}(z)dz d \sigma_{b}.\label{eq-ten5}
	\end{equation}
	By comparing Eqs.~(\ref{eq-ten1}-\ref{eq-ten5}), the expression of splitting function $d {\cal{P}}_{BA}(z)dz$ as follows
	\begin{equation}
		d {\cal{P}}_{BA}(z)dz=\frac{E_B}{E_A}\frac{M^2_{A\to{}B+C}}{(2E_B)^2(E_B+E_C-E_A)^2},
	\end{equation}
	where the masses of all particles are neglected in the calculation of the above equation. In the infinite momentum frame, the momentum of the particles are expressed in parameterized form as
	\begin{equation}
		\begin{aligned}
			&k_{A}=(P ; P, 0) ,\\
			&k_{B}=\left(z P+\frac{p_{\perp}^{2}}{2 z P} ; z P, p_{\perp}\right) ,\\
			&k_{C}=\left((1-z) P+\frac{p_{\perp}^{2}}{2(1-z) P} ;(1-z) P,-p_{\perp}\right).
		\end{aligned}
	\end{equation}
	Therefore
	\begin{align}
		& \left(2 E_{B}\right)^{2}\left(E_{A}-E_{B}-E_{c}\right)^{2}=\frac{\left(p_{\perp}^{2}\right)^{2}}{(1-z)^{2}},\\
		& \frac{d^{3} k_{C}}{(2 \pi)^{3}\left(2 E_{C}\right)} =\frac{d p_{\perp}^{2} d z}{16 \pi^{2}(1-z)}.
	\end{align}
	The splitting function is thus obtained as
	\begin{equation}
		\begin{aligned}
			d \mathcal{P}_{B A}(z) =\frac{z(1-z)}{2} \overline{\sum}_{\text{spins}} \frac{\left| V_{A \rightarrow B+C}\right|^{2} }{p_{\perp}^{2}} \frac{g^{2}}{8 \pi^{2}} \frac{d p_{\perp}^{2}}{p_{\perp}^{2}} =\frac{\alpha}{2 \pi} \frac{z(1-z)}{2} \overline{\sum}_{\text{spins}} \frac{\left| V_{A \rightarrow B+C}\right| ^{2}}{p_{\perp}^{2}} d \ln p_{\perp}^{2} .
		\end{aligned}
	\end{equation}
	
	In the following we begin to calculate the specific form of the splitting function $P_{qq}(z)$, $P_{qN}(z)$, $P_{qC}(z)$, $P_{Nq}(z)$, $P_{NC}(z)$, $P_{Cq}(z)$, $P_{CN}(z)$, $P_{CC}(z)$ under the Abelian decomposition.
	
	\begin{figure}[H]
		\centering
		\includegraphics[width=0.2\textwidth]{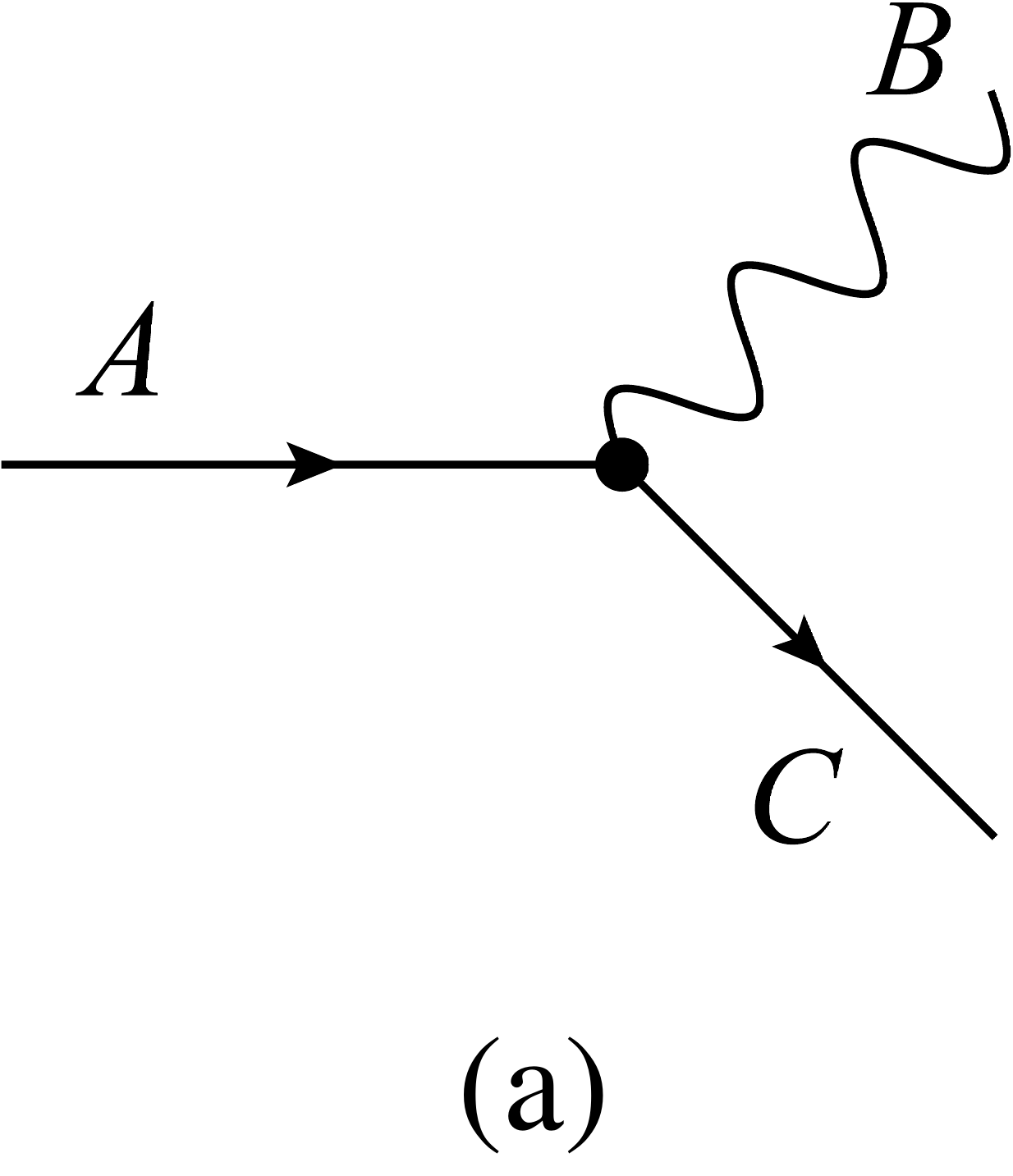}
        \hspace{1.5cm}
                \vspace{0.5cm}
        \includegraphics[width=0.2\textwidth]{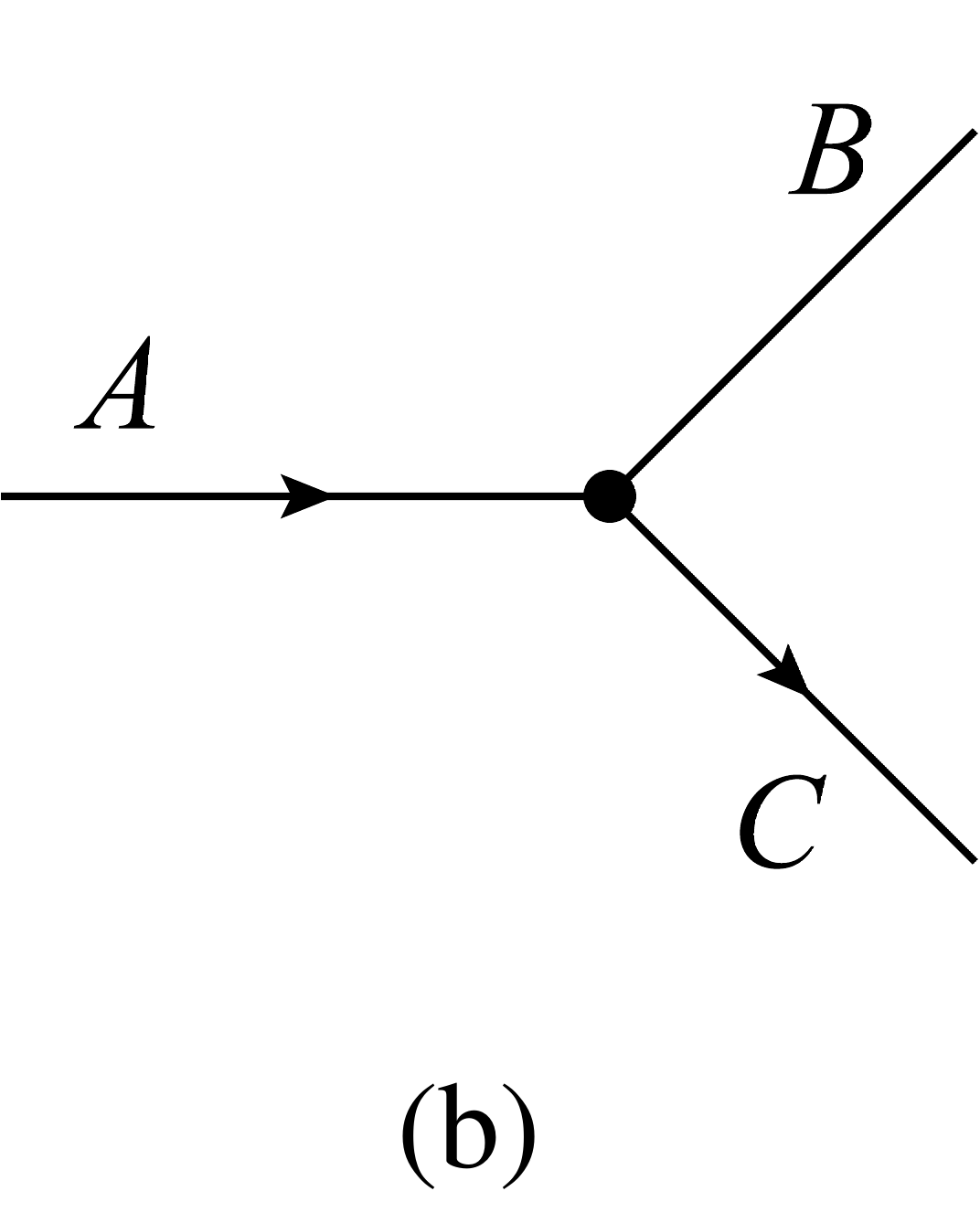}
        \hspace{1.5cm}
                \vspace{0.5cm}
        \includegraphics[width=0.2\textwidth]{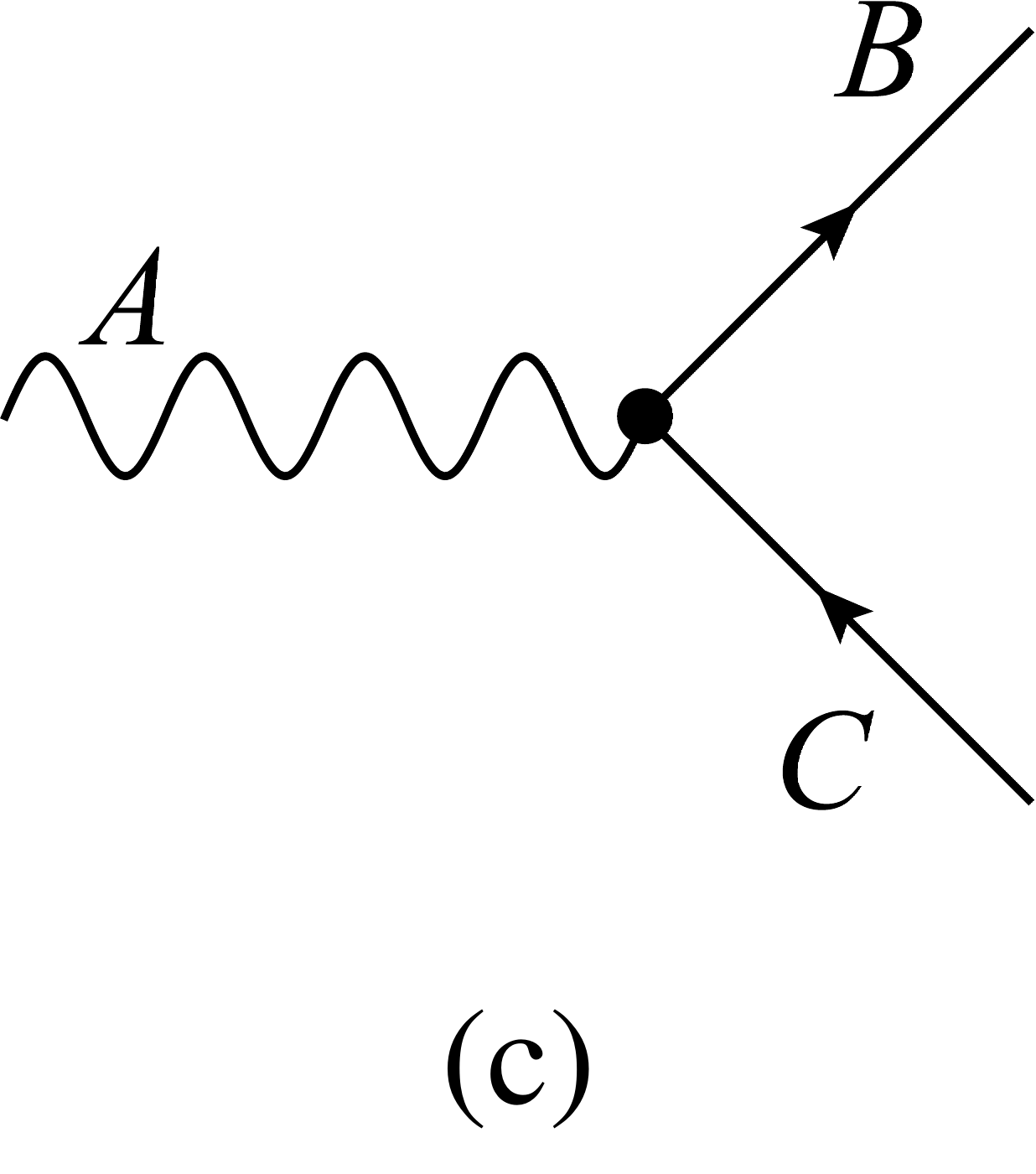}
        \vspace{0.5cm}
        \includegraphics[width=0.2\textwidth]{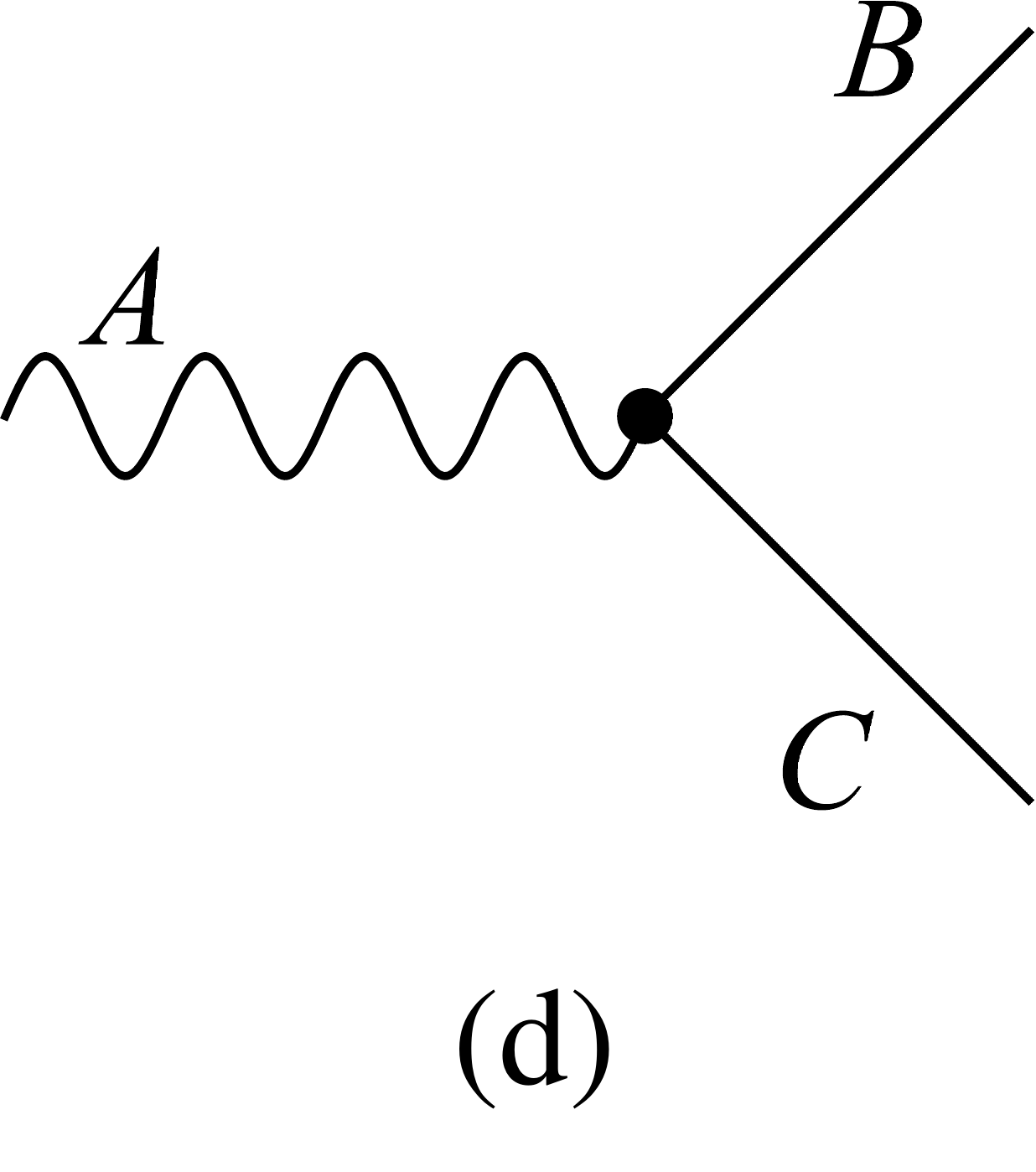}
        \hspace{2.5cm}
		\includegraphics[width=0.2\textwidth]{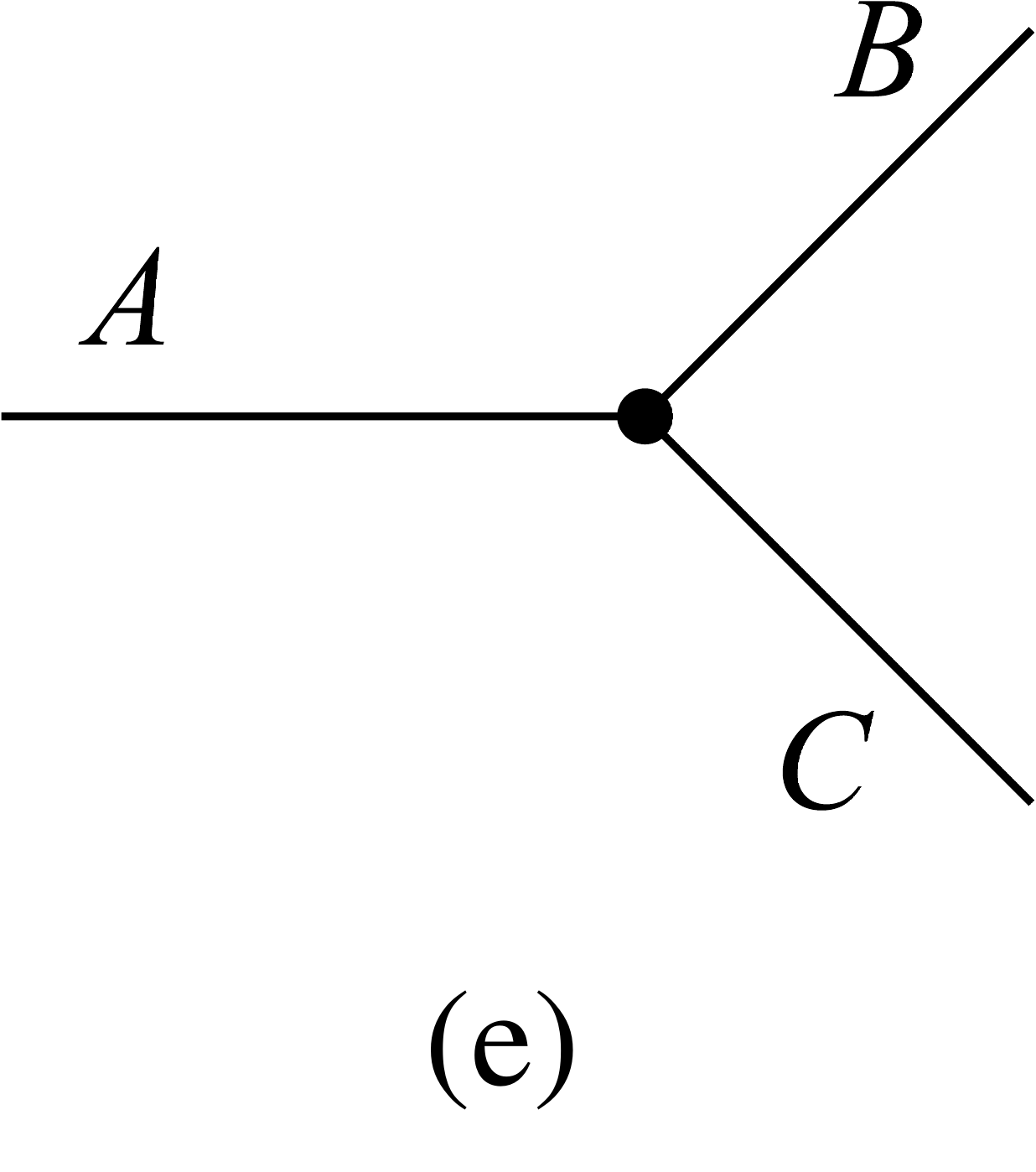}
		\caption{The vertices: (a) the quark-quark-neuron vertex (QQN), (b) the quark-quark-chromon vertex (QQC), (c) the neuron-quark-quark vertex (NQQ), (d) the neuron-chromon-chromon vertex (NCC), (e) the three-chromon vertex (CCC). }
		\label{fig-qqN}
	\end{figure}
	
	Let us first consider the quark-quark-neuron vertex (QQN) as shown in Fig.~\ref{fig-qqN}(a). The splitting function $P_{Nq}(z)$ is obtained in the form as follows
	\begin{equation}
		\overline{\sum}_{\text{spins}}\left| V_{q\to Nq}\right|^{2}=\frac{1}{2}\hat{C}_{2}(R)\operatorname{Tr}\left(\not k_{C}\gamma_{\mu}\not k_{A}\gamma_{\nu}\right)\overline{\sum}_{\text{pol}}\epsilon^{*\mu}\epsilon^{\nu},
	\end{equation}
where the coefficient $1/2$ is obtained by averaging over the initial state quark spins. Summing over the color space final states and averaging over the initial states, the $\hat{C}_{2}(R)$ is given by
	\begin{equation}
		\hat{C}_{2}(R)=\frac{1}{N}\mathrm{Tr}(\frac{t^{\hat{a}}}{2}\frac{t^{\hat{a}}}{2})=\frac{1}{3},\,\hat{a}= \,3,8.
	\end{equation}
	 Then
	\begin{equation}
		\sum_{\text{spins}}\epsilon^{*i}\epsilon^j
  \rightarrow \delta^{ij}-\frac{k^{i}_{B}k^{j}_{B}}{k_{B}^{2}}\quad (i,j=1,2,3).
	\end{equation}
	Thus
	\begin{equation}
		\overline{\sum}_{\text{spins}}\left| V_{q\to Nq}\right|^{2}=\frac{p_{\perp}^{2}}{z(1-z)}\frac{1+(1-z)^2}{z}\hat{C}_{2}(R).
	\end{equation}
	Since $z<1$, the expression of the splitting function $P_{Nq}(z)$ is obtained as
	\begin{equation}
		P_{Nq}(z)=\hat{C}_{2}(R)\frac{1+(1-z)^2}{z}.
	\end{equation}
	According to the symmetry relation $P_{qq}^{N}(z)=P_{Nq}(1-z)$, the form of the splitting function $P_{qq}^{N}(z)$ is directly obtained without additional redundant calculations
	\begin{equation}
		P_{qq}^{N}(z)= \hat{C}_{2}(R)\frac{1+z^2}{(1-z)}.
	\end{equation}
	
	The quark-quark-chromon vertex (QQC) is shown in Fig.~\ref{fig-qqN}(b). Similarly,
	\begin{equation}
		\overline{\sum}_{\text{spins}}\left| V_{q\to Cq}\right|^{2}=\frac{1}{2}C_{2}(R)\operatorname{Tr}\left(\not k_{C}\gamma_{\mu}\not k_{B}\gamma_{\nu}\right)\overline{\sum}_{\text{pol}}\epsilon^{*\mu}\epsilon^{\nu},
	\end{equation}
 where $ C_{2}(R)=1$.
	Thus,
	\begin{equation}
		\overline{\sum}_{\text{spins}}\left| V_{q\to Cq}\right|^{2}=\frac{p_{\perp}^{2}}{z(1-z)}\frac{1+(1-z)^2}{z}C_{2}(R).
	\end{equation}
	The the form of splitting function $P_{Cq}(z)$ is obtained as
	\begin{equation}
		P_{Cq}(z)=C_{2}(R)\frac{1+(1-z)^2}{z}.
	\end{equation}
	According to the similar symmetry relation $P_{qq}^{C}(z)=P_{Cq}(1-z)$, the form of the splitting function $P_{qq}^{C}(z)$ is directly obtained as
	\begin{equation}
		P_{qq}^{C}(z)= C_{2}(R)\frac{1+z^2}{(1-z)}.
	\end{equation}
	
The neuron-quark-quark vertex (NQQ) is shown in Fig.~\ref{fig-qqN}(c), which leads to the splitting function of $P_{qN}(z)$. Since $P_{qN}(z)$ is proportional to the probability density of finding a quark (or antiquark) of any color for a given flavor within the color-averaged gluon, we have
	\begin{equation}
		\overline{\sum}_{\text{spins}}\left| V_{N\to q\bar{q}}\right|^{2}=\frac{g^2}{4}\operatorname{Tr}\left(\not k_{C}\gamma_{\mu}\not k_{A}\gamma_{\nu}\right)\overline{\sum}_{\text{pol}}\epsilon^{*\mu}\epsilon^{\nu}.
	\end{equation}
	Thus
	\begin{equation}
		\overline{\sum}_{\text{spins}}\left| V_{N\to q\bar{q}}\right|^{2}=g^2 p_{\perp}^{2}\left(\frac{1-z}{z}+\frac{z}{1-z}\right).
	\end{equation}
	The splitting function $P_{qN}(z)$ is given by
	\begin{equation}
		P_{qN}(z)=\frac{1}{2}\left[z+(1-z)^{2}\right],
	\end{equation}
	According to the symmetry relation, thus we have
	\begin{equation}
		P_{qN}(z)= P_{\bar{q}N}(1-z)=P_{qN}(1-z)=\frac{1}{2}\left[z+(1-z)^{2}\right].
	\end{equation}

	The neuron-chromon-chromon vertex (NCC) ia shown in Fig.~\ref{fig-qqN}(d). So, we have
	\begin{equation}
		V_{N\to CC}=
                gf^{a\hat{b}c}\left(g_{\mu\nu}\left(-k_A-k_B\right)_\lambda+g_{\nu\lambda}\left(-k_B-k_C\right)_\mu + g_{\lambda\mu}\left(-k_C-k_A\right)_\nu \right).
	\end{equation}
	Thus
	\begin{equation}
		\overline{\sum}_{\text{spins}}\left| V_{N\to CC}\right|^{2}=12 g^2 \frac{p_{\perp}^{2}}{z(1-z)}\left[z(1-z)+\frac{1-z}{z}+\frac{z}{1-z}\right].
	\end{equation}
	The corresponding splitting function $P_{CN}(z)$ is given by
	\begin{equation}
		P_{CN}(z)=6\left[z(1-z)+\frac{1-z}{z}+\frac{z}{1-z}\right],
	\end{equation}
        Similarly,
	\begin{equation}
		P_{NC}(z)=2\left[z(1-z)+\frac{1-z}{z}+\frac{z}{1-z}\right],
	\end{equation}
		\begin{equation}
		P^N_{CC}(z)=2\left[z(1-z)+\frac{1-z}{z}+\frac{z}{1-z}\right].
	\end{equation}
	
	By analyzing the three-chromon vertex (CCC) shown in Fig.~\ref{fig-qqN}(e), the expression of splitting function $P_{CC}(z)$ is obtained through
	\begin{equation}
		V_{C\to CC}=- gf^{abc}\left(g_{\mu\nu}\left(-k_A-k_B\right)_\lambda+g_{\nu\lambda}\left(-k_B-k_C\right)_\mu + g_{\lambda\mu}\left(-k_C-k_A\right)_\nu \right).
	\end{equation}
	thus we obtain	the splitting function $P_{CC}(z)$ in Fig.~\ref{fig-qqN}(e)
	\begin{equation}
		P^C_{CC}(z)=2\left[z(1-z)+\frac{1-z}{z}+\frac{z}{1-z}\right].
	\end{equation}
	
By dealing with the singularity, similar with the original DGLAP splitting functions~\cite{Altarelli:1977zs}, the corresponding specific expressions of all splitting functions is obtained as follows
 \begin{equation}
	\begin{aligned}
		&\begin{aligned}P_{qq}=P^N_{qq}+P^C_{qq}=\frac{4}{3}\left(\frac{1+z^2}{(1-z)_+} +\frac{3}{2}\delta(1-z)\right),\end{aligned} \\
		&\begin{aligned}P_{qN}=P_{qC}=\frac{1}{2}\left( z^2+(1-z)^2\right),\end{aligned} \\
		&\begin{aligned} P_{Nq}=\frac{1}{3}\frac{1+(1-z)^2}{z},\end{aligned} \\
		&\begin{aligned}P_{Cq}=\frac{3}{3}\frac{1+(1-z)^2}{z},\end{aligned} \\
		&\begin{aligned}P_{CN}=6\left( z(1-z)+\frac{1-z}{z}+\frac{z}{(1-z)_+}+(\frac{11}{12}-\frac{n_f}{18})\delta(1-z) \right),\end{aligned}\\
    &\begin{aligned}P_{CC}=P^N_{CC}+P^C_{CC}=4\left( z(1-z)+\frac{1-z}{z}+\frac{z}{(1-z)_+}+(\frac{11}{12}-\frac{n_f}{18})\delta(1-z)\right),\end{aligned}\\
    &\begin{aligned}P_{NC}=2\left( z(1-z)+\frac{1-z}{z}+\frac{z}{(1-z)_+}+(\frac{11}{12}-\frac{n_f}{18})\delta(1-z)\right).\end{aligned}
	\end{aligned}
	\label{eq:Pij}
\end{equation}

Thus, we have completed the derivation of the splitting functions under the Abelian decomposition of QCD. According to the obtained the splitting functions, it is easy to find that the sum of the contributions of neurons and chromons to the evolution is equal to the contribution of gluons. This result is in accordance with our expectation, since the Abelian decomposition of QCD does not make any changes to original QCD, but merely decomposes it, resulting in the full evolution unchanged.

With the splitting functions in Eq. (\ref{eq:Pij}), we present the new DGLAP evolution equations involving neuron and chromon. Let $C(x,t)$ represent the chromon density and $N(x,t)$ represent the neuron density. The new DGLAP equations after the Abelian decomposition are as follows:
	\begin{equation}
		\begin{aligned}
			&\frac{dq^i(x,t)}{dt} =\frac{\alpha(t)}{2\pi}\int_x^1\frac{dy}y\left[\sum_{j=1}^{2f}q^j(y,t)P_{q^iq^j}(\frac xy)+N(y,t)P_{q^iN}(\frac xy)+C(y,t)P_{q^iC}(\frac xy)\right],  \\
			&\frac{dN(x,t)}{dt} =\frac{\alpha(t)}{2\pi}\int_x^1\frac{dy}y\left[\sum_{j=1}^{2f}q^j(y,t)P_{Nq^j}(\frac xy)+C(y,t)P_{NC}(\frac xy)\right],  \\
			&\frac{dC(x,t)}{dt} =\frac{\alpha(t)}{2\pi}\int_x^1\frac{dy}y\left[\sum_{j=1}^{2f}q^j(y,t)P_{Cq^j}(\frac xy)+N(y,t)P_{CN}(\frac xy)+C(y,t)P_{CC}(\frac xy)\right].
		\end{aligned}
		\label{eq:NewDGLAP}
	\end{equation}
	In Eq. (\ref{eq:NewDGLAP}), it is evident that additional splitting functions are introduced for the interactions between neurons and chromons, as well as between neurons and quarks, and between chromons and quarks, compared to Eq. (\ref{eq:DGLAP}).

\begin{figure*}[b]
	\centering
 	\includegraphics[width=0.6\textwidth]{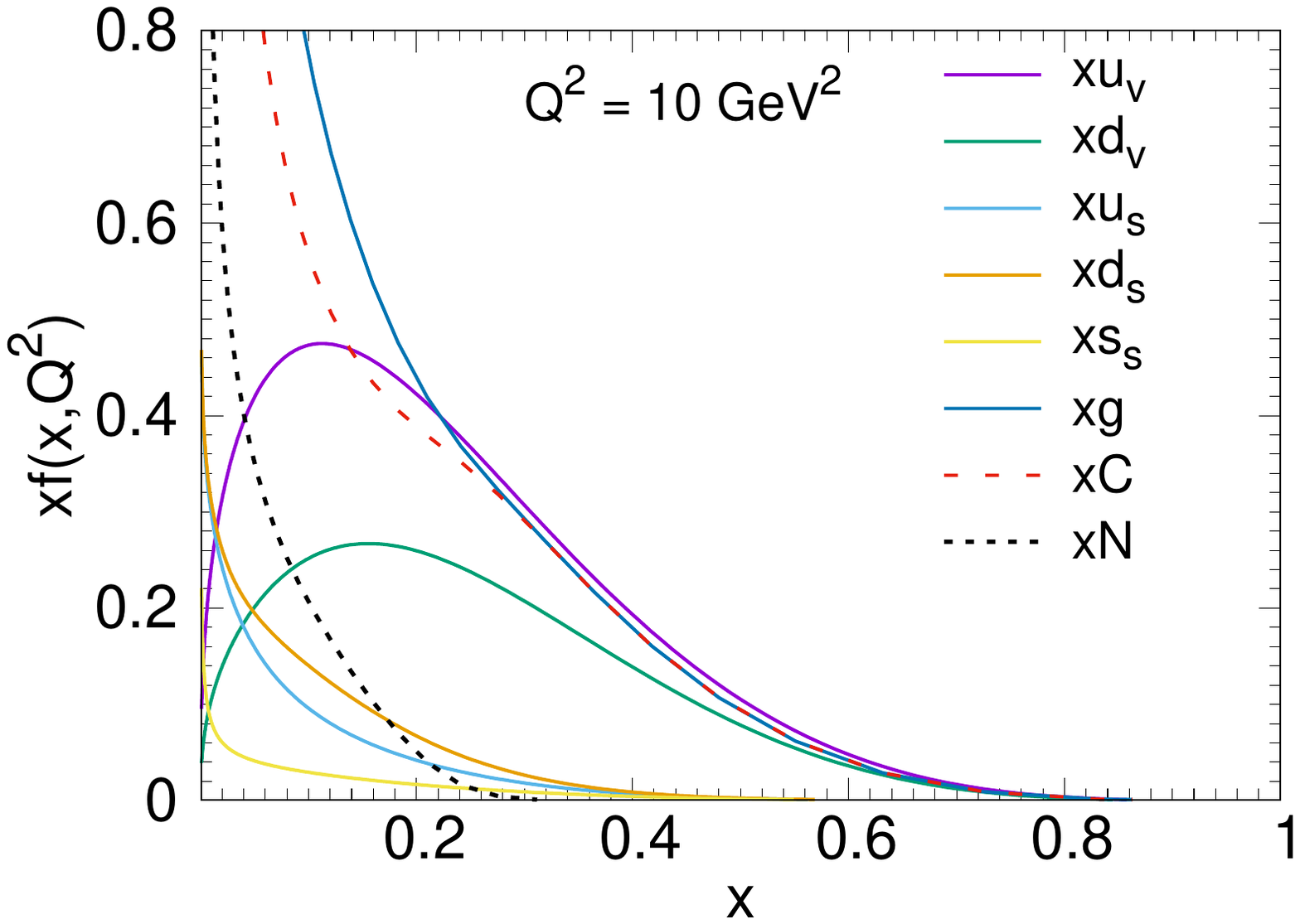}
	\label{fig:xf5}
	\caption{The parton distributions at $Q^2$ evolving to 10 GeV$^2$ with initial ratio $R_i=9:1$. $xu_v$ and $xd_v$ denote the distributions of up valence quark and down valence quark, respectively. $xu_s$, $xd_s$ and $xs_s$ denote the distributions of up sea quark, down sea quark and strange sea quark, respectively. $xg$, $xC$ and $xN$ denote the distributions of total gluon (chromon + neuron), chromon and neuron, respectively.}
	\label{fig:xq-xN-xC-xG}
\end{figure*}
It is evident that, due to the fact that the decomposition itself does not alter the gauge symmetry of QCD, the total contributions of the two types of gluons is consistent with the gluon contribution in the original QCD, as required by physical reality. To verify this idea, we numerically solve the Abelian-decomposed QCD evolution equations. As an initial attempt, we provide the proton parton distribution at a specific evolution starting point and evolve it using our new evolution equations. We adopt the initial gluon parameterization scheme from Ref. \cite{Gluck:1998xa} and set the initial distributions of chromon and neuron to be $xC(x,Q_0^2)=0.9 xg(x,Q_0^2)$ and $xN(x,Q_0^2)=0.1 xg(x,Q_0^2)$, respectively, with $Q_0^2=0.26$ GeV$^2$. That is the initial ratio $R_i$ of the number of chromons to neurons is approximately 3:1. The initial gluon distribution is $xg(x,Q_0^2)=17.47x^{1.6}(1-x)^{3.8}$. By using Eqs. (\ref{eq:Pij}) and (\ref{eq:NewDGLAP}), the evolution results at $Q^2 ~= 10$ GeV$^2$ are shown in Fig. \ref{fig:xq-xN-xC-xG}. We can see that the total contributions from the two types of gluons are consistent with the gluon contribution in the original QCD evolution results. The Abelian decomposition does not influence the quark distribution.

\begin{figure*}[b]
	\centering
 	\includegraphics[width=0.45\textwidth]{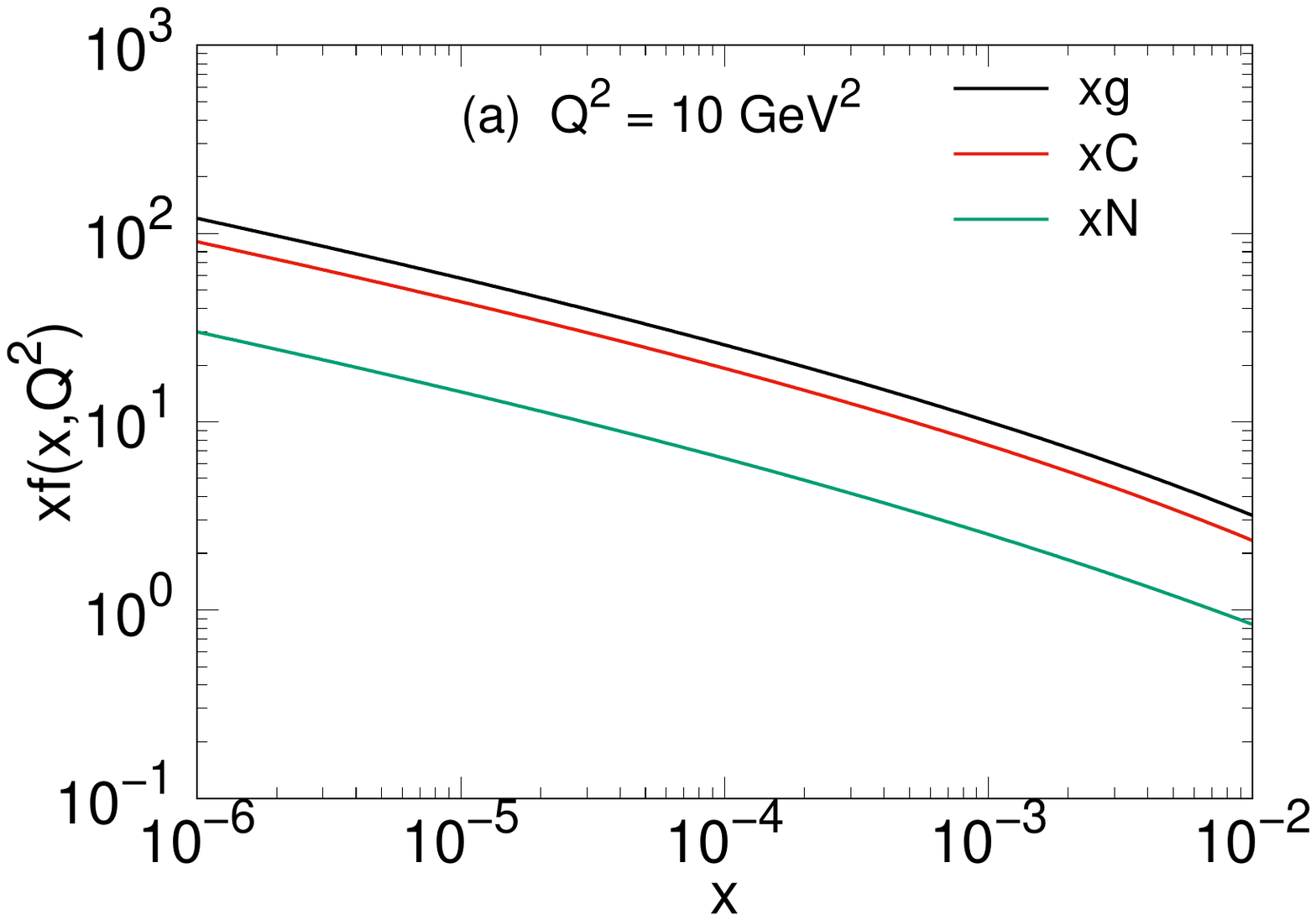}
	\label{fig:xg10}
	\includegraphics[width=0.45\textwidth]{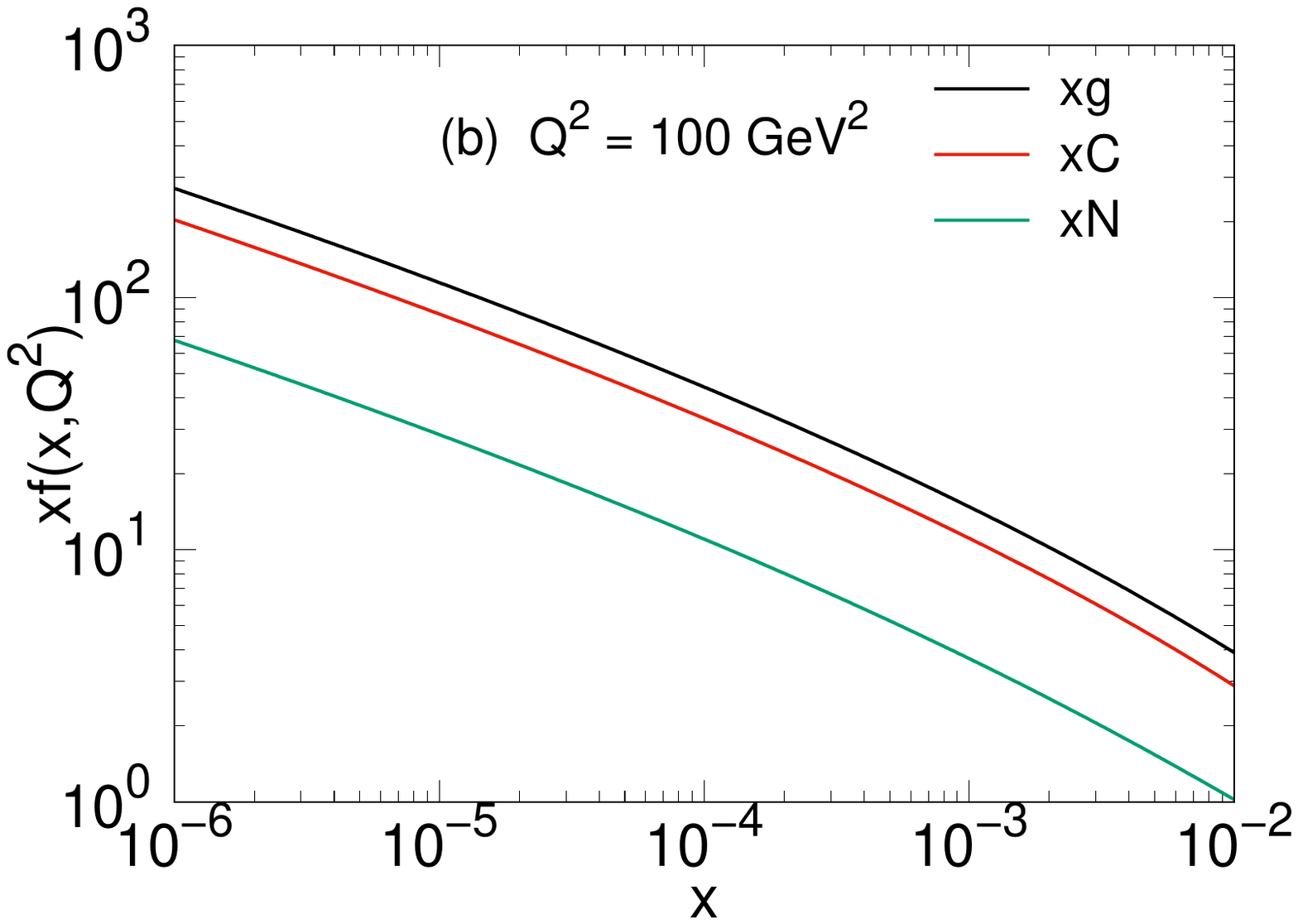}
	\label{fig:xg100}
	\includegraphics[width=0.45\textwidth]{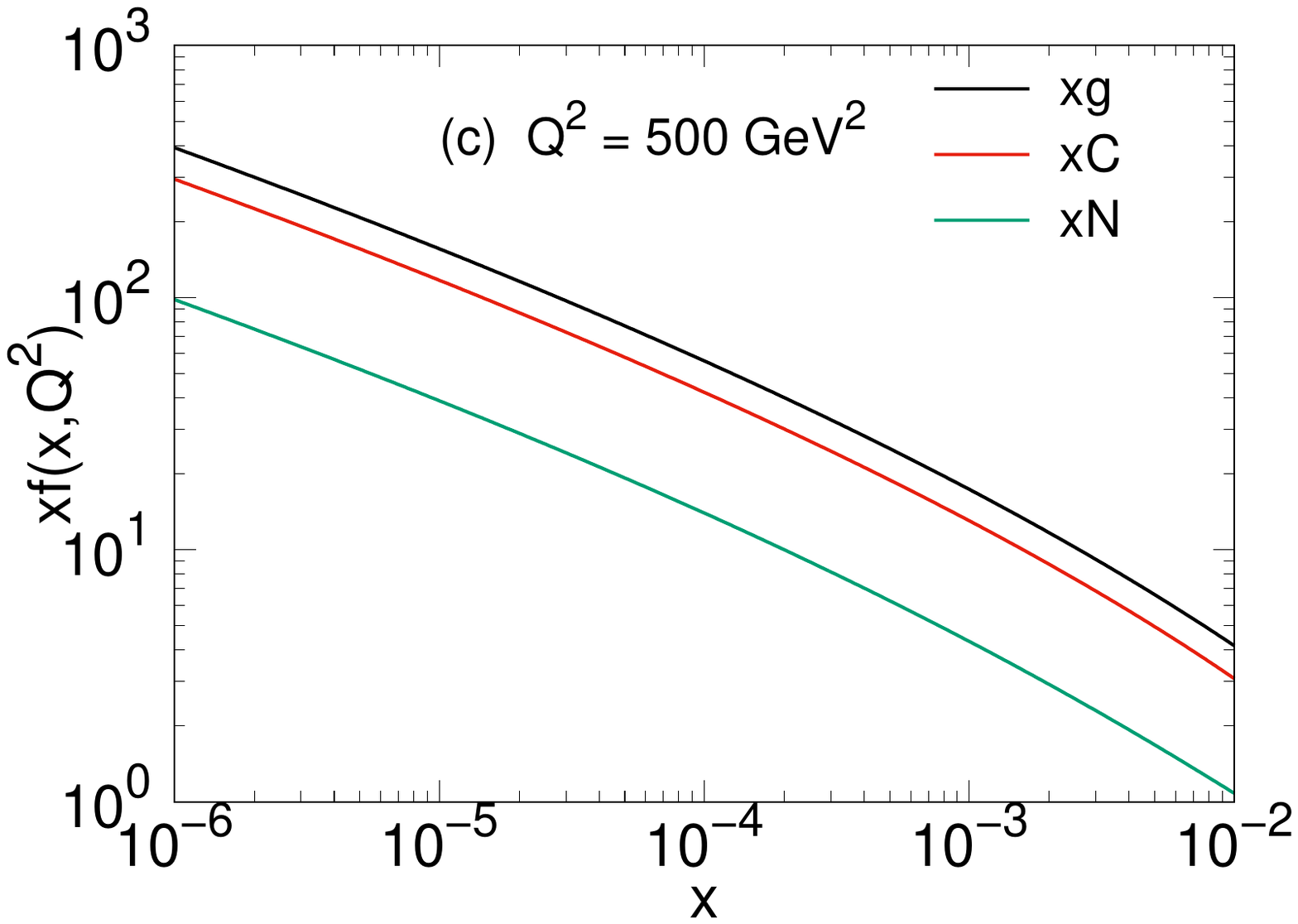}
	\label{fig:xg500}
	\includegraphics[width=0.45\textwidth]{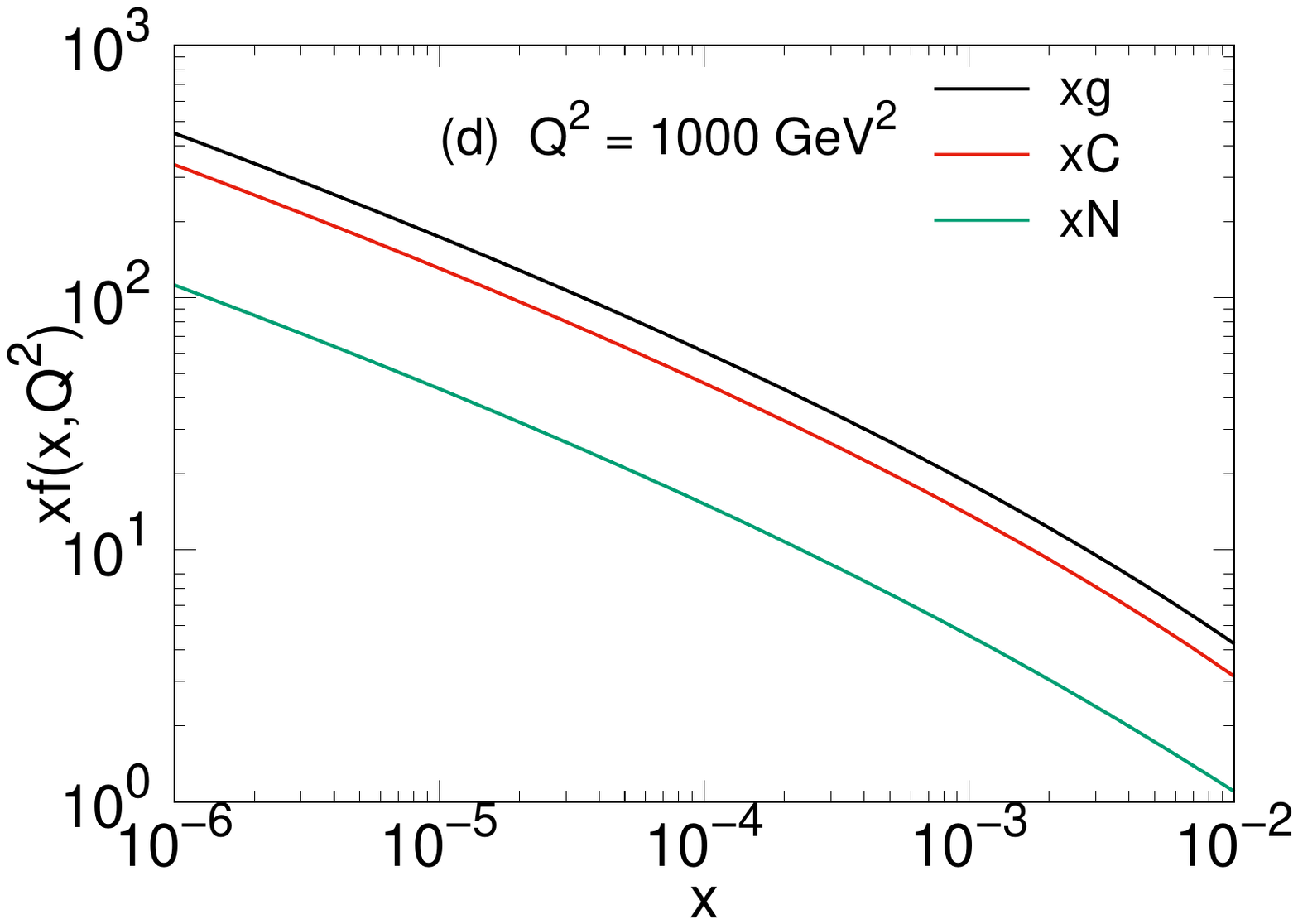}
	\label{fig:xg1000}
	\caption{The gluon distributions at $Q^2$ evolving to 10, 100, 500, 1000 GeV$^2$. The red lines represent the chromon distribution, the green lines represent the neuron distribution, and the black lines represent the total contribution of the two types of gluons, i.e., the gluon distribution in original QCD evolution results.}
	\label{fig:xN-xC-xG}
\end{figure*}
\begin{figure*}[h]
	\centering
 	\includegraphics[width=0.45\textwidth]{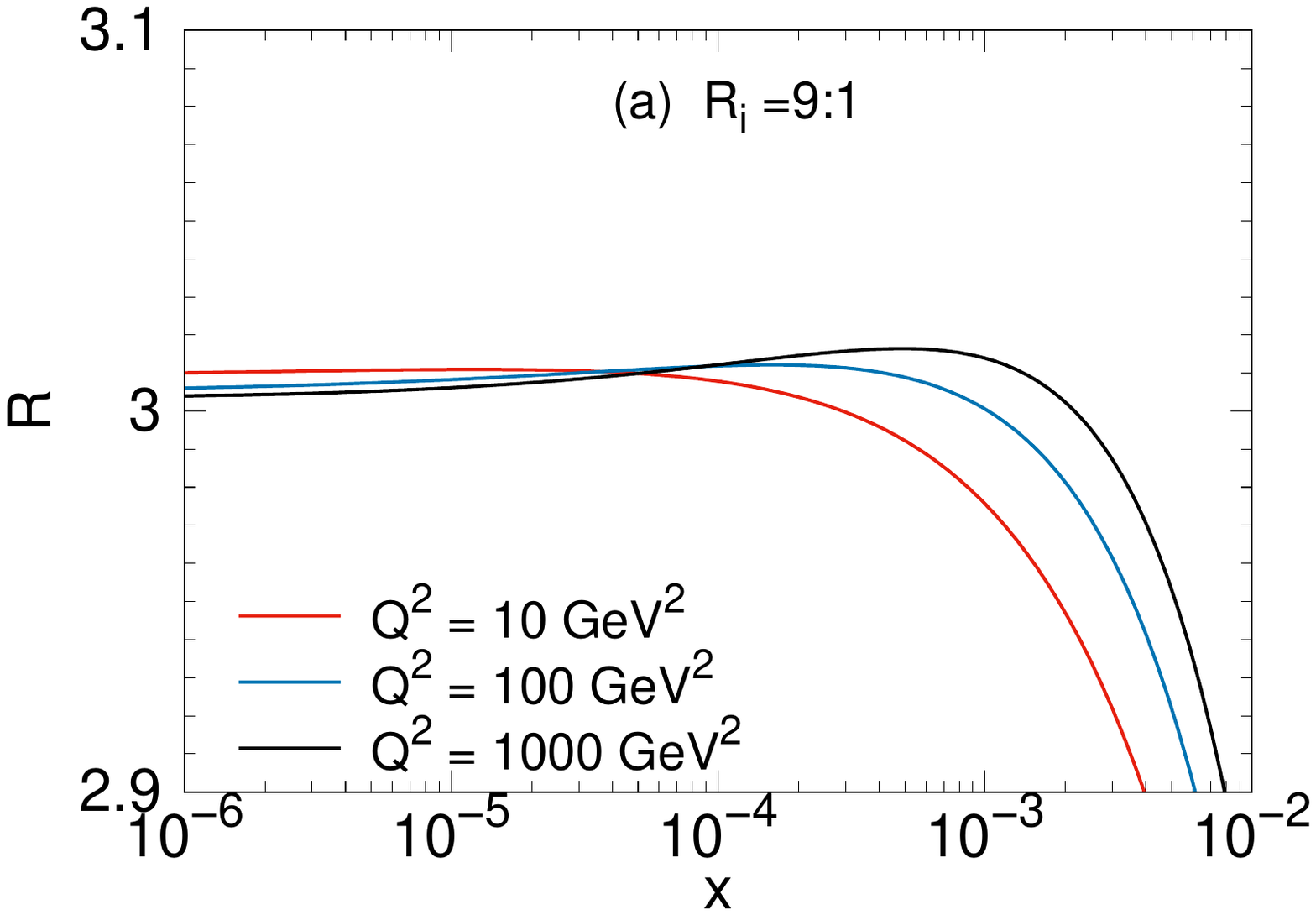}
	\label{fig:r0d9}
	\includegraphics[width=0.45\textwidth]{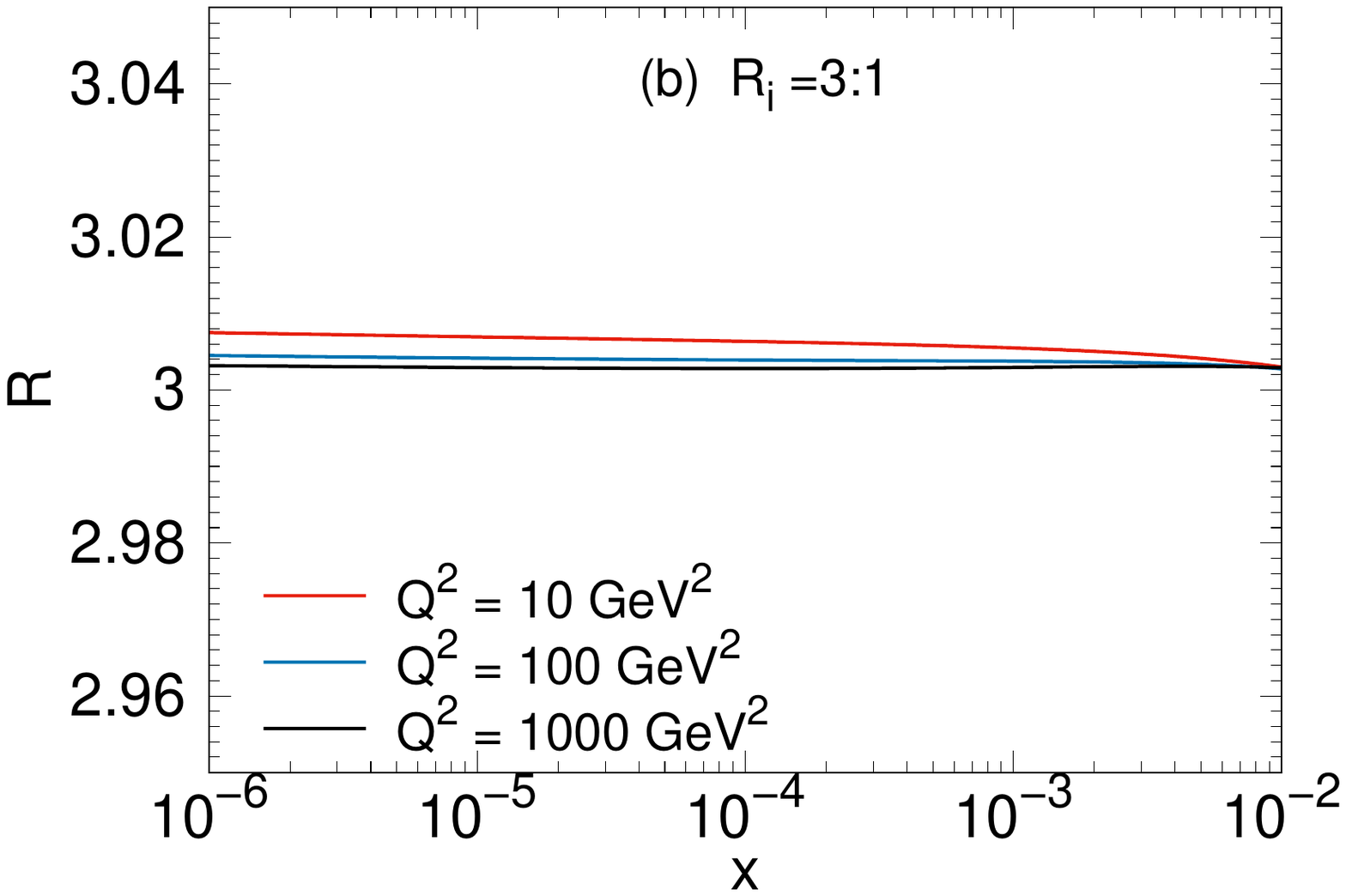}
	\label{fig:r0d75}
 	\includegraphics[width=0.45\textwidth]{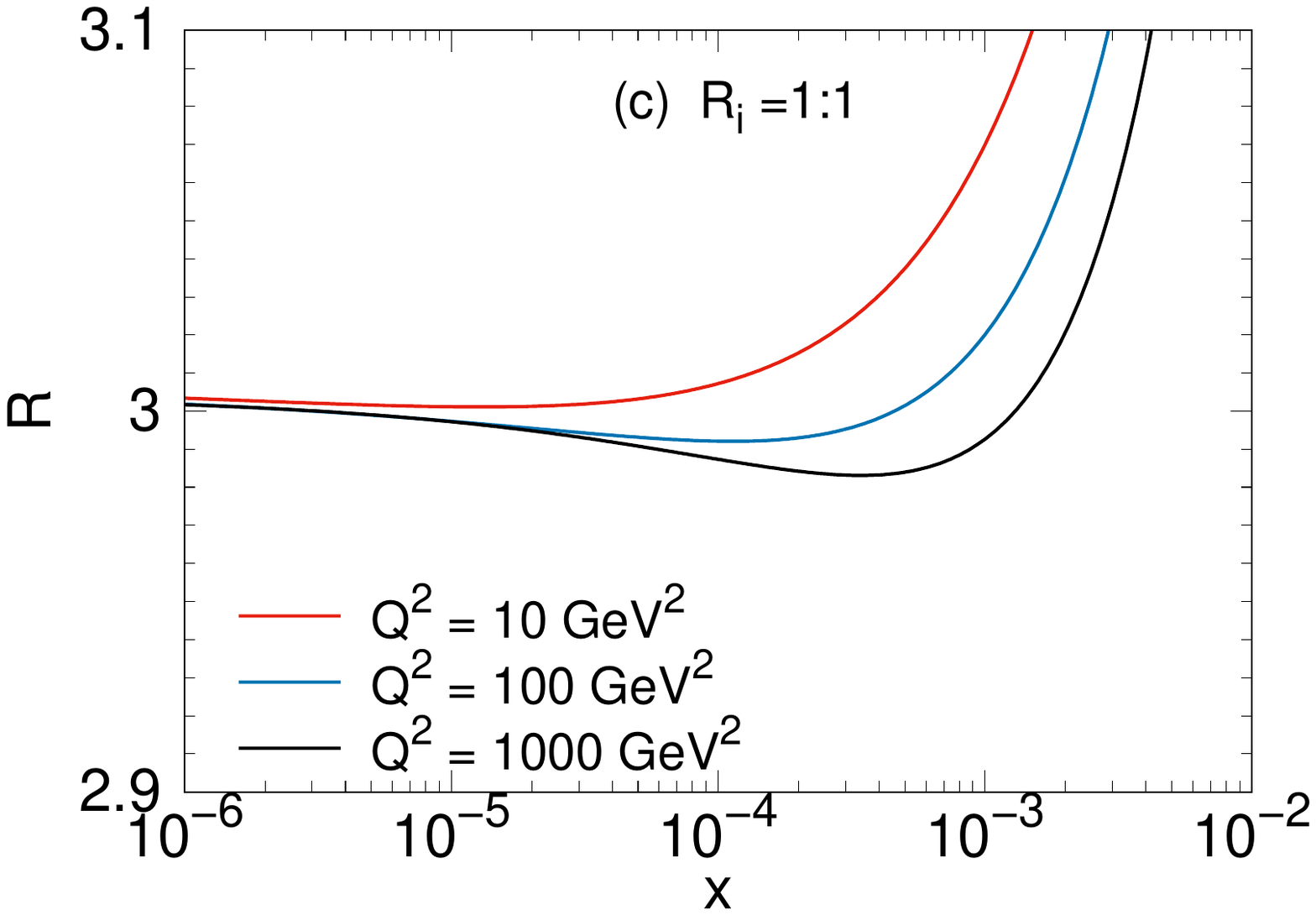}
	\label{fig:r0d75}
	\includegraphics[width=0.45\textwidth]{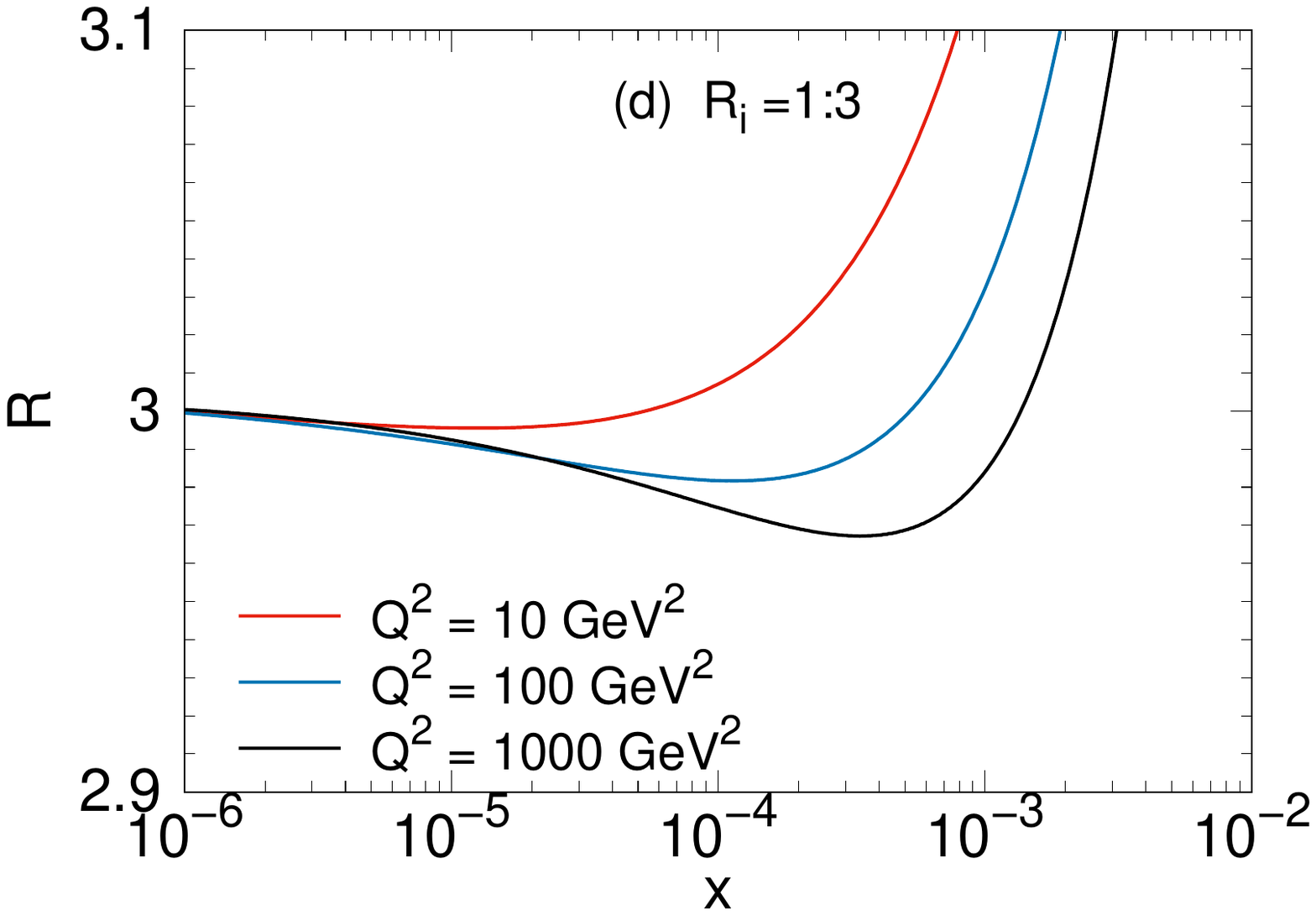}
	\label{fig:r0d25}
	\includegraphics[width=0.45\textwidth]{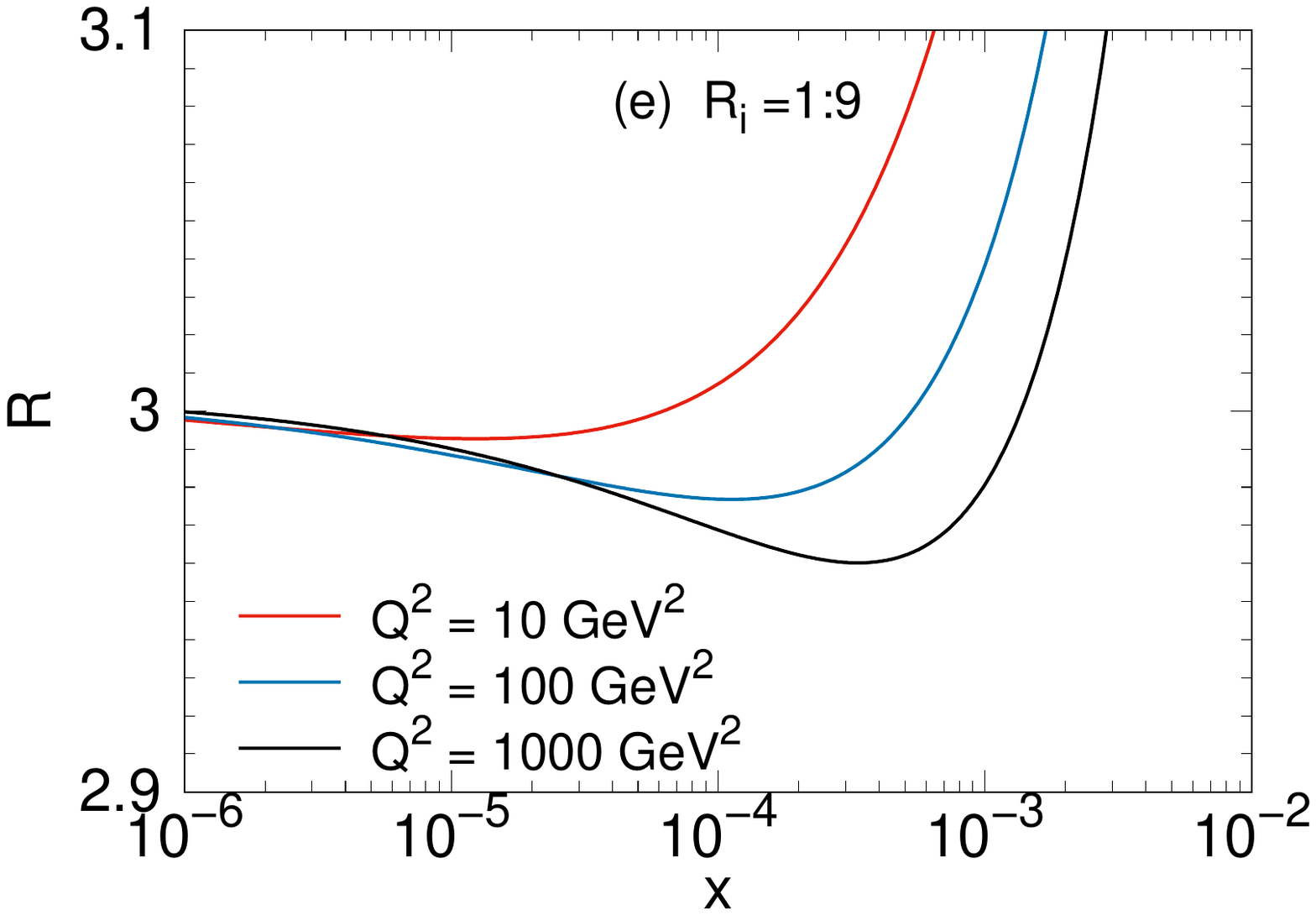}
	\label{fig:r0d1}
	\caption{The ratios of the number of chromons to neurons at different $Q^2$. The initial ratios are $9:1$ (a), $3:1$ (b), $1:1$ (c), $1:3$ (d), and $1:9$ (e), respectively.}
	\label{fig:R}
\end{figure*}

For better perspective of the gluon distributions, we present the results of the contributions from the two types of gluons as well as the total gluon distribution at different scales $Q^2$ as shown in Fig. \ref{fig:xN-xC-xG}. The evolution results of the two types of gluons reveal that the combined distribution contributions of neurons and chromons are consistent with the complete QCD gluon distribution. Moreover, we also investigate the gluon distributions with different ratios of the number of chromons to neurons at the initial gluon distributions. Table \ref{tab:R} shows ratios of the number of chromons to neurons at $x=10^{-4}$ with different initial ratios $R_i$. We find that regardless of the choice of the evolution starting point, the ratio of the number of chromons to neurons is approximately 3:1 in small-$x$ region when it evolutes to high $Q^2$. Figure \ref{fig:R} shows the ratio results at different $Q^2$ with different initial ratios. This is a surprising phenomenon, indicating the presence of both color-neutral and colored gluons within the proton, with their combined gluon contributions to the proton being approximately in a 3:1 ratio. We have emphasized the roles of these two types of gluons in previous sections. On one hand, the color-neutral neurons provide binding interactions, akin to the photons in QED, as they cannot change the color of quarks. On the other hand, the colored chromons maintain the full color gauge symmetry, enabling the exchange of color charge in the scattering processes with quarks. Additionally, the topological gauge monopole, which do not participate in dynamics, provide confinement, confining the two types of gluons and quarks within an asymptotically free space.
\begin{table}[h!]
  \begin{center}
  \caption{The ratios of the number of chromons to neurons at $x=10^{-4}$ with different $R_i$.}
  \begin{tabular}{cccccccccc}
  \hline
   & & $~R_i = 9:1~$ & $~R_i = 3:1~$ & $~R_i = 1:1~$ & $~R_i = 1:3~$ & $~R_i = 1:9~$ \\
  \hline
  &$Q^2 = 10$ GeV$^2$ & 3.008   & 3.006  & 3.007  & 3.007   & 3.007  \\
  \hline
    &$Q^2 = 100$ GeV$^2$ & 3.012   & 3.004  & 2.992  & 2.982   & 2.977  \\
  \hline
    &$Q^2 = 1000$ GeV$^2$ & 3.012   & 3.003  & 2.987  & 2.975   & 2.969  \\
  \hline
    \end{tabular}%
  \label{tab:R}
  \end{center}
\end{table}

It is necessary to investigate the contributions of neurons and chromons \cite{Cho:2022rib}, particularly their relative proportions, in the gluon jets produced in high-energy collision experiments. This is motivated by the different ways in which various gluons produce jets. From the decomposition results of three- and four-gluon vertices, it can be observed that the jets produced by neurons resemble those produced by photons, involving the emission of (anti-)chromon pairs. However, apart from emitting (anti-)chromon pairs, chromons also emit neuron pairs, individual chromon emissions, and individual neuron emissions. This suggests that even a qualitative analysis reveals differences in the proportions of final state jets. The simulation work is valuable, and considering the dynamical evolution of different gluons will provide effective assistance to the simulation process.

\section{Conclusion and outlook}
\label{sec:summary}
In summary, we have derived a new set of evolution equations for the classical DGLAP evolution equation based on the CDG's Abelian decomposition of QCD. The new equations incorporate the dynamical behavior of the two types of gluons introduced in the CDG decomposition. These two types of gluons play distinct roles within hadrons, with the color-neutral neurons behaving akin to photons in QED, binding other constituent particles and addressing the question of how quarks are bound within hadrons in the quark model. The colored gluons maintain the full color gauge symmetry, acting as color sources and exchanging color with the colored constituents. The only non-dynamical topological monopole provides confinement, as discussed in the literature. The new evolution equations describe the dynamical behavior of the two types of gluons separately. We find that the ratio of chromons to neurons remains approximately 3:1 throughout the evolution process. This presents a more surprising result, and the underlying physics requires exploration from the evolution equations themselves and the Abelian decomposition of QCD. The new evolution equations derived are not in contradiction with the classical DGLAP scheme, as the Abelian decomposition does not alter the internal structure of QCD.

The dynamical behavior of the two types of gluons needs experimental validation, and we hope that the new evolution equations can provide new insights for calculating the cross sections of gluon jet production in high-energy collisions \cite{ATLAS:2013uet,ATLAS:2014hvo,ATLAS:2014vax,ATLAS:2016vxz,ATLAS:2017bje,CMS:2014sip,CMS:2015ebl,CMS:2021iwu,Gallicchio:2011xq,Larkoski:2013eya,Bhattacherjee:2015psa,FerreiradeLima:2016gcz,Metodiev:2018ftz,Davighi:2017hok,Gras:2017jty,Komiske:2018vkc,Larkoski:2019nwj} and for Monte Carlo simulations of related physical processes.

\begin{acknowledgments}
We are very grateful for the valuable discussion and communication with K. Yang, P. Zhang, W. Zhu, and L. Zou. This work is supported by the Strategic Priority Research Program of Chinese Academy of Sciences under Grant No.XDB34030301; Guangdong Major Project of Basic and Applied Basic Research No.2020B0301030008; Guizhou Provincial Basic Research Program (Natural Science) under Grant No.QKHJC-ZK[2023]YB027; Education Department of Guizhou Province under Grant No.QJJ[2022]016.
\end{acknowledgments}

\bibliographystyle{apsrev4-1}
\bibliography{refs}

\newpage
\appendix
\setcounter{figure}{0}
\renewcommand{\thefigure}{\Alph{section}.\arabic{figure}}
\renewcommand{\thetable}{\Alph{section}.\arabic{table}}

\end{document}